\documentclass[11pt,onecolumn,draftcls]{IEEEtran} \newcommand{\figsize}{3.85}
\usepackage{graphicx,psfrag}
\usepackage[usenames]{color}
\usepackage[cmex10]{amsmath}
\usepackage{amsfonts,amssymb,latexsym,cite,ifthen,color}
\newboolean{SEP_FIG_CAPS}\setboolean{SEP_FIG_CAPS}{false}

\ifthenelse{\boolean{SEP_FIG_CAPS}}
{% if separate figures/tables/captions...
 \newcommand{\putFrag}[4]{\begin{figure}[p]
                            \centering
                            #4
			    \includegraphics[width=#3in]{figures/#1.eps}
            		    \caption{}
     			    \label{fig:#1}
                          \end{figure}
                          \clearpage}
 \newcommand{\putTable}[3]{\begin{table}[p]
  			    \centering
		            #3
            		    \caption{}
     			    \label{tab:#1}
			  \end{table}
			  \clearpage}
 \newcommand{\capFrag}[2]{\noindent Fig.~\ref{fig:#1}. #2 \medskip\\}
 \newcommand{\capTable}[2]{\noindent Tab.~\ref{tab:#1}. #2 \medskip\\}
}
{% elseif not separate figures/tables/captions...
 \newcommand{\putFrag}[4]{\begin{figure}[t]
                            \centering
                            #4
			    \includegraphics[width=#3in]{figures/#1.eps}
            		    \caption{#2}
           		    \label{fig:#1}
                          \end{figure} }
 \newcommand{\putTable}[3]{\begin{table}[t]
  			    \centering
		            #3
     			    \caption{#2}
     			    \label{tab:#1}
			  \end{table} }
 \newcommand{\capFrag}[2]{}
 \newcommand{\capTable}[2]{}
}

 \newcommand{\defn}{\triangleq}

 \newcommand{\tvec}[1]{\ensuremath{\Tilde{\boldsymbol{#1}}}}
 
 \newcommand{\hvec}[1]{\ensuremath{\Hat{\boldsymbol{#1}}}}
    
 \renewcommand{\vec}[1]{\ensuremath{\boldsymbol{#1}}}
 \newcommand{\mat}[1]{\ensuremath{\begin{bmatrix}#1\end{bmatrix}}}

 \newcommand{\norm}[1]{\ensuremath{\| #1 \|}}
 \newcommand{\mc}[1]{\ensuremath{\mathcal{#1}}}

 \newcommand{\Real}{{\mathbb{R}}}
 \newcommand{\Complex}{{\mathbb{C}}}

 \newcommand{\conv}{\star}

 \newcommand{\tran}{^\textsf{T}}
 \newcommand{\herm}{^\textsf{H}}
 \newcommand{\of}[1]{^{(#1)}}

 \DeclareMathOperator{\real}{Re}

 \DeclareMathOperator{\E}{E}
 \DeclareMathOperator{\var}{var}
 \DeclareMathOperator{\cov}{Cov}

 \DeclareMathOperator{\Diag}{{\mc D}}

 \renewcommand{\eqref}[1]{(\ref{eq:#1})}
 
 \newcommand{\Figref}[1]{Figure~\ref{fig:#1}}
 \newcommand{\figref}[1]{Fig.~\ref{fig:#1}}
 \newcommand{\tabref}[1]{Table~\ref{tab:#1}}
 \newcommand{\secref}[1]{Section~\ref{sec:#1}}
 
 \newcommand{\appref}[1]{Appendix~\ref{app:#1}}

 \newcounter{comment}[section]
 
 \newcounter{texthead}[section]
 
 \renewenvironment{itemize}
   {\begin{list}{\labelitemi}{\topsep 0.05in \itemsep 0in}}{\end{list}}
\newcommand{\giv}{\,|\,}

\newcommand{\const}{\mathbb{S}}

\newcommand{\Np}{N_\textsf{p}}
\newcommand{\Nd}{N_\textsf{d}}
\newcommand{\Md}{M_\textsf{d}}
\newcommand{\Mc}{M_\textsf{c}}
\newcommand{\Mi}{M_\textsf{i}}
\newcommand{\Mt}{M_\textsf{t}}
\newcommand{\pt}{_\textsf{pt}}

\newcommand{\BER}{\textsf{BER}}
\newcommand{\NMSE}{\textsf{NMSE}}

\newcommand{\gt}{g_{\text{\sf t}}}
\newcommand{\gr}{g_{\text{\sf r}}}
\newcommand{\Lpre}{L_{\text{\sf pre}}}

\newcommand{\inp}{_{\textsf{in},j}}
\newcommand{\out}{_{\textsf{out},i}}
\newcommand{\pri}{^\text{\sf apri}}

\newcommand{\ext}{^\text{\sf ext}}
\newcommand{\lmmse}{_\text{\sf lmmse}}
\newcommand{\lasso}{_\text{\sf lasso}}

\begin{document}
\setlength{\arraycolsep}{0.8mm}
 \title{A Message-Passing Receiver for 
 	BICM-OFDM over Unknown Clustered-Sparse Channels}
	 \author{Philip~Schniter$^*$%and Andreas~F.~Molisch
         \thanks{Please direct all correspondence to 
                Prof.\ Philip Schniter,
                Dept. ECE, 
		The Ohio State University,
		2015 Neil Ave., Columbus OH 43210,
                e-mail: schniter@ece.osu.edu,
                phone 614.247.6488, fax 614.292.7596.}
	 %\thanks{A. Molisch is with the Dept.\ of EES at
	 %		   the University of Southern California,
	 %		   Los Angeles, CA 90089.
	 %		   (Email: molisch@usc.edu)}
         \thanks{This work has been supported in part by 
	 	NSF grant CCF-1018368 and 
		DARPA/ONR grant N66001-10-1-4090, and 
		an allocation of computing time from the Ohio Supercomputer Center.}
	}
 \date{\today}
 \maketitle

\begin{abstract}
We propose a factor-graph-based approach to joint 
channel-estimation-and-decoding (JCED) of bit-interleaved coded 
orthogonal frequency division multiplexing (BICM-OFDM).
In contrast to existing designs, ours is capable of exploiting not only 
sparsity in sampled channel taps but also clustering among the large taps, 
behaviors which are known to manifest at larger communication bandwidths.
In order to exploit these channel-tap structures, we adopt a two-state 
Gaussian mixture prior in conjunction with a Markov model on the hidden state. 
For loopy belief propagation, we exploit a ``generalized approximate
message passing'' (GAMP) algorithm recently developed in the context of 
compressed sensing, and show that it can be successfully coupled with 
soft-input soft-output decoding, as well as hidden Markov inference, 
through the standard sum-product framework.
For $N$ subcarriers and any channel length $L<N$,
the resulting JCED-GAMP scheme has a computational complexity of only 
$\mc{O}(N\log_2 N + N |\const|)$, where $|\const|$ is the constellation size.
Numerical experiments using IEEE~802.15.4a channels
show that our scheme yields BER performance within 1 dB of
the known-channel bound and 
3-4 dB better than soft equalization based on LMMSE and LASSO.
\end{abstract}

%%%%%%%%%%%%%%%%%%%%%%%%%%%%%%%%%%%%%%%%%%%%%%%%%%%%%%%%%%%%%%%%%%%%%%%%%%%%%
\section{Introduction} 				\label{sec:intro}

When designing a digital communications receiver, it is common to model the 
effects of multipath propagation in discrete time using a convolutive 
linear channel that, in the slow-fading scenario, can be characterized 
by a fixed impulse response $\{x_j\}_{j=0}^{L-1}$ over 
the duration of one codeword.
When the communication bandwidth is sufficiently low, the ``taps''
$\{x_j\}_{j=0}^{L-1}$ are well modeled as independent complex Gaussian 
random variables, resulting in the ``uncorrelated Rayleigh-fading'' and 
``uncorrelated Rician-fading'' models that have dominated the wireless 
communications literature for many decades \cite{Molisch:Book:05}.
For receiver design, the Gaussian tap assumption is very convenient 
because the optimal estimation scheme is well known to be linear 
\cite{Poor:Book:94}.
As the communication bandwidth increases, however, the channel taps
are no longer well-modeled as Gaussian nor independent.
Rather, they tend to be heavy-tailed or ``sparse'' in that only 
a few values in $\{x_j\}_{j=0}^{L-1}$ have significant amplitude
\cite{Cramer:TAP:02,Preisig:JAcSA:04,Molisch:TVT:05,Czink:TWC:07}. %Hashemi:PROC:93
Moreover, groups of large taps are often clustered together in lag $j$.
These behaviors are both a blessing and a curse: a blessing because, of all
%the performance of optimal channel estimation improves (i.e., 
tap distributions, the independent Gaussian one is most detrimental to 
capacity \cite{Medard:TIT:00}, but a curse because optimal channel estimation 
becomes non-linear and thus receiver design becomes more complicated.

Recently, there have been many attempts to apply breakthrough 
non-linear estimation techniques from the field of ``compressive sensing'' 
\cite{Mar:SPM:08} (e.g., LASSO \cite{Tibshirani:JRSSb:96,Chen:JSC:98}) 
to the wireless channel estimation problem.
We refer to this approach as ``compressed channel sensing'' (CCS), 
after the recent comprehensive overview \cite{Bajwa:PROC:10}.
The CCS literature generally takes a \emph{decoupled} approach to 
the problem of channel estimation and data decoding, in that pilot-symbol 
knowledge is first exploited for sparse-channel estimation, after which
the channel estimate is used for data decoding.
%Therein, the hope is that channel sparsity can be leveraged for 
%pilot-rate reduction, with the end goal of increased spectral efficiency.
However, this decoupled approach is known to be suboptimal 
\cite{Kannu:arXiv:10}.

The considerations above motivate a \emph{joint} approach to
structured-sparse-channel-estimation and decoding that offers both 
near-optimal decoding performance and low implementation complexity.
In this paper, we propose exactly such a scheme. 
In particular, we focus on orthogonal frequency-division 
multiplexing (OFDM) with bit-interleaved coded modulation (BICM),
and propose a novel factor-graph-based receiver that leverages recent
results in ``generalized approximate message passing'' (GAMP)
\cite{Rangan:10b}, 
soft-input/soft-output (SISO) decoding \cite{MacKay:Book:03},
and structured-sparse estimation \cite{Schniter:CISS:10}.
Our receiver assumes a clustered-sparse channel-tap prior constructed using 
a two-state Gaussian mixture with a Markov model on the hidden
tap state.
The scheme that we propose has only $\mc{O}(N\log_2 N\!+\!N|\const|)$ 
complexity, where $N$ denotes the number of subcarriers and $|\const|$ 
denotes the constellation size,
facilitating large values of $N$ and channel length $L\!<\!N$
(e.g., we use $N\!=\!1024$ and $L\!=\!256$ for our numerical results).
For rich non-line-of-sight (NLOS) channels generated according to the
IEEE~802.15.4a standard \cite{Molisch:802.15.4a}, 
%and while operating at high spectral efficiencies, 
our numerical experiments show bit error rate ($\BER$) 
performance within $1$ dB of the known-channel bound
and $3$--$4$ dB better than soft equalization based on LMMSE and LASSO.

We now place our work in the context of existing factor-graph designs.
Factor-graph based joint channel-estimation and decoding (JCED) 
was proposed more than a decade ago
(see, e.g., the early overview \cite{Worthen:TIT:01}).
To calculate the messages passed among the nodes of the factor graph,
first instincts suggest to apply the standard 
``sum-product algorithm'' (SPA) 
\cite{Pearl:Book:88,Loeliger:Proc:07,Kschischang:TIT:01}.
Exact SPA on the JCED factor graph is computationally infeasible,
however, and so it must be approximated.
For this, there are many options, since many well-known iterative inference 
algorithms can themselves be recognized as SPA approximations, e.g.,
the expectation-maximization (EM) algorithm %\cite{Dempster:JRSS:77}
\cite{Dauwels:ISIT:05},
particle filtering \cite{Dauwels:ISIT:06},
variational (or ``mean-field'') techniques \cite{Dauwels:ISIT:07},
and even steepest descent \cite{Dauwels:ITW:05}.
Moreover, because the JCED factor graph is loopy, 
even non-approximate SPA is not guaranteed to yield the correct 
output distributions, because exact inference is NP hard \cite{Cooper:AI:90}.
%Thus, given the plethora of approximation techniques and uncertainty
%about exact SPA itself, 
It is perhaps not surprising that, amidst this uncertainty about exact 
SPA and its ``best'' approximation, a number of different factor-graph 
approaches to JCED over frequency-selective channels have been proposed
(e.g., \cite{Novak:ICASSP:09,Liu:PIMRC:09,Knievel:ICC:10,Kirkelund:GLOBE:10}).
%
% Novak, Matz, Hlawatsch - OFDM/IDMA with unknown-length ISI 
% Liu, Brunel, Boutros - single-carrier with ISI channel
% Knievel, Shi, Hoeher, Auer - MIMO-OFDM with time/freq correlation
% Kirkelund, Manchon, Chritensen, Riegler, Fleury - MIMO-OFDM 
%

Our approach differs from existing factor-graph JCED designs in that
it uses
1) a sparse (i.e., non-Gaussian) channel-tap prior, 
%non-Gaussian) whereas, in all of the existing BP-based JCED work that
%we are aware, the channel coefficients are modeled as Gaussian, and
2) a clustered (i.e., non-independent) channel-tap prior, and 
3) a state-of-the-art SPA approximation known as ``generalized approximate 
message passing'' (GAMP), which has been shown to 
admit rigorous analysis
as $N,L\!\rightarrow\!\infty$ \cite{Rangan:10b}.
In fact, we conjecture that the success of our method is due in large
part to the principled approximations used within GAMP.
We also note that, although we focus on the case of clustered-sparse channels,
our approach could be applied to non-sparse (i.e., Gaussian) 
or non-clustered (i.e., independent) channel-taps or, e.g.,
non-sparse channels with unknown length $L$ \cite{Novak:ICASSP:09},
with minor modifications of our assumed channel prior.

Finally, we mention that this work is an evolution of our earlier work 
\cite{Schniter:ASIL:10,Schniter:PHYCOM:11} that was 
limited to an exactly sparse channel, that did not exploit clustering,
and that was based on the ``relaxed belief propagation'' (RBP) algorithm
\cite{Rangan:10v2}, which has higher implementation complexity
than GAMP.
For example, the JCED scheme from
\cite{Schniter:ASIL:10,Schniter:PHYCOM:11} has complexity 
$\mc{O}(NL\!+\!N|\const|)$, which grows with the channel length $L$.

Our paper is organized as follows.
In \secref{model} we detail our assumptions on the OFDM system 
and the channel prior, and provide an illustrative example of
clustered-sparse behavior with the IEEE~802.15.4a channel model.
In \secref{jced} we detail our GAMP-based JCED approach,
in \secref{sims} we report the results of our simulation study, and
in \secref{conc} we conclude.

%\footnote{
  Throughout the paper, we use the following notation.
  $\Real$ denotes the field of reals and 
  $\Complex$ the complex field. 
  $(\cdot)^*$ denotes conjugate and
  $\real(\cdot)$ extracts the real part.
  %$\Int$ the set of integers.
  Furthermore,
  $\delta(\tau)$ denotes the Dirac delta waveform while
  $\{\delta_n\}_{n=-\infty}^\infty$ denotes the Kronecker delta sequence.
  %the sinc function is denoted $\sinc(x)=\sin(x)/x$,
  Also,
  $\langle j\rangle_N$ denotes $j$-modulo-$N$,
  $\conv$ convolution, and
  $\propto$ denotes equality up to a scaling.
  We use
  boldface capital letters like $\vec{B}$ to denote matrices and 
  boldface small letters like $\vec{b}$ to denote vectors.
  %$\cir(\vec{b})$ denotes the circulant matrix with first column $\vec{b}$,
  $\vec{I}$ denotes the identity matrix,
  $\vec{1}$ denotes the vector of ones, and 
  $\Diag(\vec{b})$ constructs a diagonal matrix from the vector $\vec{b}$.
  %$\diag(\vec{B})$ the diagonal matrix with the same diagonal terms 
  %       as matrix $\vec{B}$, and
  %We use $[\vec{B}]_{m,n}$ to denote the element in the $m^{th}$ row
  %and $n^{th}$ column of $\vec{B}$, where row/column indices begin
  %with zero.
  For matrices and vectors,
  $(\cdot)\tran$ denotes transpose and
  $(\cdot)\herm$ denotes conjugate transpose. 
  %$\norm{\cdot}_F$ denotes the Frobenius norm,
  %$\norm{\cdot}$ the $\ell_2$ norm, 
  %$(\cdot)^+$ the Moore-Penrose pseudo-inverse, and
  %$\odot$ element-wise multiplication. 
  %
  When $x_j$ is a realization of random variable $X_j$, we write
  $x_j\!\sim\! X_j$ and use
  $\E_{X_j}\{x_j\}$ to denote the mean,
  $\var_{X_j}\{x_j\}$ the variance,
  %covariance by $\cov\{\vec{b},\vec{c}\} \defn \E\{\vec{b}\vec{c}^H\}
  %       -\E\{\vec{b}\}\E\{\vec{c}^H\}$. 
  $p_{X_j}(x_j)$ the pdf, and 
  $p_{X_j|D_j}(x_j\giv d_j)$ the pdf conditioned on the event $D_j\!=\!d_j$.
  Sometimes we omit the subscript when there is no danger of confusion, 
  	yielding, e.g., $\E\{x_j\}$, $\var\{x_j\}$, $p(x_j)$ and $p(x_j\giv d_j)$.
  $\mc{CN}(x;\hat{x},\nu^x)\!\defn\!(\pi\nu^x)^{-1}\exp(-|x-\hat{x}|^2/\nu^x)$ 
     	denotes the circular Gaussian pdf with mean $\hat{x}$ and variance $\nu^x$.
  In fact, we often use $(\hat{v}_j,\nu_j^v)$ when referring to the 
	mean and variance of $V_j$.
  For a random vector $\vec{x}$, we use $\cov(\vec{x})$ to denote the
  covariance matrix.
%}

%%%%%%%%%%%%%%%%%%%%%%%%%%%%%%%%%%%%%%%%%%%%%%%%%%%%%%%%%%%%%%%%%%%%%%%%%%%%%
\section{System Model} 				\label{sec:model}

\subsection{The BICM-OFDM model}		\label{sec:OFDM}

We consider an OFDM system with $N$ subcarriers, each modulated by a 
QAM symbol from a $2^M$-ary unit-energy constellation $\const$.
Of the $N$ subcarriers, $\Np$ are dedicated as pilots,\footnote{
  For our GAMP decoder, we recommend $\Np\!=\!0$; see \secref{sims}.}
and the remaining $\Nd\!\defn\! N\!-\!\Np$ 
are used to transmit a total of $\Mt$ training bits 
and $\Md\! \defn\! \Nd M \!-\! \Mt$ coded/interleaved data bits.
The data bits are generated by encoding $\Mi$ information bits using a
rate-$R$ coder, interleaving them, and partitioning the resulting 
$\Mc\!\defn\!\Mi/R$ 
bits among an integer number $Q\!\defn\!\Mc/\Md$ of OFDM symbols.
We note that the resulting scheme has a spectral efficiency of 
$\eta\!\defn\! \Md R/N$ information bits per channel use (bpcu).

In the sequel, 
we use $s\of{k}\!\in\!\const$ for $k\!\in\!\{1,\dots,2^M\}$ to denote the 
$k^{th}$ element of the QAM constellation, and 
$\vec{c}\of{k}\!\defn\! [c_{1}\of{k},\dots,c_{M}\of{k}]\tran$ to denote the 
corresponding bits as defined by the symbol mapping.
Likewise, 
we use $s_i[q]\!\in\!\const$ for the QAM symbol transmitted 
on the $i^{th}$ subcarrier of the $q^{th}$ OFDM symbol and 
$\vec{c}_i[q] \!\defn\! [c_{i,1}[q],\dots,c_{i,M}[q]]\tran$ for the 
coded/interleaved bits corresponding to that symbol.
We use $\vec{c}[q] \!\defn\! [\vec{c}_0[q],\dots,\vec{c}_{N-1}[q]]\tran$
to denote the coded/interleaved bits in the $q^{th}$ OFDM symbol 
and $\vec{c} \!\defn\! [\vec{c}[1],\dots,\vec{c}[Q]]\tran$ to denote 
the entire (interleaved) codeword.
The elements of $\vec{c}$ that are apriori known as pilot or training bits 
will be referred to as $\vec{c}\pt$.
The remainder of $\vec{c}$ is determined from the information
bits $\vec{b}\!\defn\![b_1,\dots,b_{\Mi}]\tran$ by coding/interleaving.

To modulate the $q^{th}$ OFDM symbol, an $N$-point inverse discrete Fourier 
transform (DFT) $\vec{\Phi}\herm$ is applied to the QAM sequence
$\vec{s}[q]\!=\![s_0[q],\dots,s_{N-1}[q]]\tran$, yielding the time-domain 
sequence $\vec{\Phi}\herm\vec{s}[q] \!=\!  \vec{a}[q] \!=\!  
[a_0[q],\dots,a_{N-1}[q]]\tran $.
The OFDM waveform $a(t)$ is then constructed using
$L$-cyclic-prefixed versions of $\{a_{j}[q]\}$ and the transmission pulse 
$\gt(\tau)$:
\begin{eqnarray}
  a(t) &=& \sum_{q=1}^Q \sum_{j=-L}^{N-1} a_{\langle j\rangle_N}\![q]\,
	\gt\big(t-jT-q(N+L)T\big),
\end{eqnarray}
with $T$ denoting the baud interval (in seconds) and $L<N$.
%$\alpha\!\in\![0,1]$ the SRRC parameter.

The waveform $a(t)$ propagates through a noisy channel
with an impulse response $h(\tau)$ that is supported on the interval
$[\tau_{\min},\tau_{\max}]$, resulting in the receiver input waveform
\begin{eqnarray}
  r(t) &=& w(t) + \int_{\tau_{\min}}^{\tau_{\max}} h(\tau) a(t-\tau) d\tau,
\end{eqnarray}
where $w(t)$ is a Gaussian noise process with flat power spectral density $N_o$.
We note that a time-invariant channel is assumed for simplicity.
The receiver samples $r(t)$ through the reception pulse $\gr(\tau)$, obtaining 
\begin{eqnarray}
  r_j[q] &=& \int r(t) \, \gr\big(jT +q(N+L)T-t\big) dt, 
  %\quad j=0,\dots,N-1,
\end{eqnarray}
and applies an $N$-DFT $\vec{\Phi}$ to each time-domain sequence 
$\vec{r}[q]\!=\![r_0[q],\dots,r_{N-1}[q]]\tran$, yielding the frequency-domain 
sequences $\vec{\Phi}\vec{r}[q] \!=\! \vec{y}[q] \!=\! [y_0[q],\dots,y_{N-1}[q]]\tran$ 
for $q=1\dots Q$.

Defining the pulse-shaped channel response
$x(\tau) \!\defn\! (\gr \conv h\conv\gt)(\tau)$, 
it is well known (e.g., \cite{Cimini:TCOM:85}) that,
when the support of $x(\tau)$ is contained within the interval $[0,LT)$,
the frequency domain observation on the $i^{th}$ subcarrier can be written as
%$\vec{y}[q] = \Diag(\vec{s}[q]) \vec{\Phi} \vec{x}[q] + \vec{w}[q]$, i.e., 
\begin{eqnarray}
  y_i[q]
  &=& s_i[q] z_i[q] + w_i[q] ,				\label{eq:yi}
\end{eqnarray}
where $z_i[q]\in\Complex$ is the $i^{th}$ subcarrier's gain and 
$\{w_i[q]\}$ are Gaussian noise samples.
Furthermore, defining the uniformly sampled channel ``taps''
$x_j[q]\!\defn\!x(jT\!+\!q(N\!+\!L)T)$, %for $j\in\{0,\dots,L-1\}$,
the subcarrier gains are related to these taps through the DFT:
\begin{eqnarray}
  z_i[q]
  &=& \sum_{j=0}^{L-1} \Phi_{ij} x_j[q] .		\label{eq:zi}
\end{eqnarray}
In addition, when $(\gr\conv\gt)(\tau)$ is a Nyquist pulse, 
$\{w_i[q]\}_{\forall i,q}$ are statistically independent with 
variance $\nu^w\!=\!N_o$.

To simplify the development, we assume that $Q=1$ in the sequel
(but not in the simulations), and drop the index $[q]$ for brevity.

\subsection{A clustered-sparse tap prior}		\label{sec:GM2}

Empirical studies \cite{Cramer:TAP:02,
Preisig:JAcSA:04,Molisch:TVT:05,Czink:TWC:07} %Hashemi:PROC:93
have suggested that, when the baud rate $T^{-1}$ is sufficiently large, 
the channel taps $\{x_j\}$ are ``sparse'' in that the tap distributions 
tend to be heavy tailed.
%There exist a large and growing number of publications that propose 
%ways to exploit this sparsity for channel estimation
%(see, e.g., the bibliography of \cite{Bajwa:PROC:10}),
%many of which have been inspired by the theory of ``compressive sensing''
%\cite{Mar:SPM:08}.
The same empirical studies suggest that large taps tend to be clustered
in the lag domain. 
%a natural consequence of scattering physics 
%and, to some extent, the finite width of the pulse shape 
%$(\gr\conv\gt)(\tau)$.
Furthermore, both the sparsity and clustering behaviors can be 
lag-dependent, such as when the receiver's timing-synchronization 
mechanism aligns the first strong multipath arrivals with a 
particular reference lag $j$.
A concrete example of these behaviors will be given in 
\secref{IEEE}.

Since our message-passing-based receiver design is inherently Bayesian, 
we seek a prior
%\footnote{
%  Although it is sometimes suggested that the need to assume priors on
%  the channel is a weakness of the Bayesian approach, it should be 
%  noted that ``deterministic'' sparse reconstruction schemes 
%  like LASSO also require one to choose a tuning parameter that effectively
%  trades off between the estimate's sparsity and residual variance,
%  and that strongly affects the resulting mean-squared error.} 
on the taps $\{x_j\}$ that is capable of representing this lag-dependent 
clustered sparsity.
For this purpose, we assume a two-state Gaussian mixture (GM2) 
prior,\footnote{
  The message passing algorithm described in \secref{gamp}
  can also handle non-Gaussian mixtures and/or mixtures with more
  than two terms.} 
\begin{eqnarray}
  p(x_j)
  &=& (1-\lambda_j) \mc{CN}(x_j;0,\nu^0_j) +
        \lambda_j \mc{CN}(x_j;0,\nu^1_j) ,      \label{eq:pxj}
\end{eqnarray}
where   
$\nu_j^0\!\geq\! 0$ denotes the variance while in the ``small'' state,
$\nu_j^1\!>\!\nu_j^0$ denotes the variance while in the ``big'' state,
and $\lambda_j\!\defn\! \Pr\{d_j\!=\!1\}$ denotes the prior probability of 
$x_j$ being in the ``big'' state.
Here, we use $d_j\!\in\!\{0,1\}$ to denote the hidden state, implying the 
state-conditional pdf 
$%\begin{eqnarray}
  p(x_j\giv d_j)
  = \mc{CN}(x_j;0,\nu^{d_j}_j). %\label{eq:px|dj}
$%\end{eqnarray}
%Assuming that the channel is energy-preserving with  
%an exponential delay-power profile, we have
%$\nu_j \!=\! 2^{-j/\Lhpd}/(\sum_{r=0}^{L-1}\lambda_r 2^{-r/\Lhpd})$,
%where $\Lhpd$ denotes the half-power delay.

For example, if $x_j$ was presumed to be a ``sparse'' tap, then we would
choose $\lambda_j\!\ll\! 1$ and $\nu_j^1 \!\gg\! \nu_j^0$ in \eqref{pxj}.
If, on the other hand, $x_j$ is presumed to be (non-sparse) Rayleigh-fading, 
we would choose $\lambda_j\!=\!1$ and set $\nu_j^1$ equal to the tap variance, 
noting that $\nu_j^0$ becomes inconsequential.
If $x_j$ is presumed to be Nakagami-fading or similar, we could 
fit the GM2 parameters $[\lambda_j,\nu_j^0,\nu_j^1]$ appropriately using
the EM algorithm, as described in \cite[p.~435]{Bishop:Book:07}.
The GM2 prior has been used successfully in many other 
non-Gaussian inference problems (see, e.g., \cite{Ishwaran:AS:05}), and our
premise here is that the GM2 model achieves a good balance between
fidelity and tractability when modeling channel taps as well. 
%\footnote{ 
%  The extension to Gaussian mixtures with non-zero means and more than 
%  two terms is straightforward.  Using this approach, any desired pdf can be
%  modeled with arbitrary accuracy \cite{Sorensen:AUTOM:71}.}

To capture the big-tap clustering behavior, we employ a hidden Markov 
model (HMM).
For this, we model the tap states $\{d_j\}_{j=0}^{L-1}$ 
as a Markov chain (MC) with switching probabilities 
$p_j^{01}\!\defn\!\Pr\{d_{j+1}\!=\!0\giv d_{j}\!=\!1\}$
and $p_j^{10}\!\defn\!\Pr\{d_{j+1}\!=\!1\giv d_{j}\!=\!0\}$.
Here, $p_j^{01}<0.5$ implies that the neighbors of a big $x_j$
tend to be big, and $p_j^{10}<0.5$ implies that the neighbors of
a small $x_j$ tend to be small.
We note that $\{p_j^{01},p_j^{10}\}_{j=0}^{L-1}$ must be consistent with 
$\{\lambda_j\}_{j=0}^{L-1}$ in that the following must hold
for all $j$:
\begin{eqnarray}
  \mat{\lambda_{j+1}~&~1\!-\!\lambda_{j+1}} 
  &=& \mat{\lambda_{j}~&~1\!-\!\lambda_{j}} 
  	\mat{1\!-\!p_j^{01} & p_j^{01}\\ p_j^{10} & 1\!-\!p_j^{10}}.
\end{eqnarray}

Although we allow correlation among the tap states, we assume that the
tap \emph{amplitudes} are conditionally independent, i.e., 
$p(x_{j+1},x_j\giv d_{j+1},d_j) \!=\! p(x_j\giv d_j) p(x_{j+1}\giv d_{j+1})$.
Our experiences with IEEE 802.15.4a channels (see below) suggest that this 
is a valid assumption.

We emphasize that the model parameters 
$\{\lambda_j,p_j^{01},p_j^{01},\nu_j^1,\nu_j^0\}$ are allowed to vary 
with lag $j$, 
facilitating the exploitation of apriori known lag-dependencies 
in sparsity and/or clustering.

\subsection{An illustrative example: IEEE 802.15.4a channels}		\label{sec:IEEE}

As an illustrative example of the clustered-sparse tap behavior described
above, we generated realizations of the tap vector
$\vec{x}\defn[x_0,\dots,x_{L-1}]\tran$ from channel impulse responses 
$h(\tau)$ generated according to the method specified in the IEEE~802.15.4a 
``ultra-wideband'' standard \cite{Molisch:802.15.4a}, which uses the
Saleh-Valenzuela model \cite{Saleh:JSAC:87}
\begin{eqnarray}
  h(\tau)
  &=& \sum_{c=0}^C\sum_{k=0}^K h_{k,c}  
        e^{j\phi_{k,c}} \delta(\tau-T_c-\tau_{k,c}),
\end{eqnarray}
where $C$ denotes the number of clusters, $T_c$ the delay of the $c^{th}$
cluster, $K$ the number of components per cluster, $\{\tau_{k,c}\}$ the
relative component delays, $\{h_{k,c}\}$ the component amplitudes,
and $\{\phi_{k,c}\}$ the component phases.
In particular, the 802.15.4a standard specifies the following.
\begin{itemize}
\item
The cluster arrival times are a Poisson process with rate $\Lambda$,
i.e., $p(T_c\giv T_{c-1})\!=\! \Lambda \exp(-\Lambda(T_c-T_{c-1}))$.
The initial cluster delay $T_0\!\geq\! \tau_{\min}$, as seen by the receiver,
is a function of the timing synchronization algorithm.
\item
The component arrivals are a mixture of two Poisson processes:
$p(\tau_{k,c} |\tau_{k-1,c}) 
\!=\! \beta\lambda_1 \exp(-\lambda_1(\tau_{k,c}-\tau_{k-1,c}))
\!+\! (1-\beta)\lambda_2 \exp(-\lambda_2(\tau_{k,c}-\tau_{k-1,c}))$
with $\tau_{0,c}=0$.
\item
The component energies obey
\begin{eqnarray}
  \E\{|h_{k,c}|^2\}
  &=& \frac{\exp(-T_l/\Gamma-\tau_{k,l}/\gamma)}
        {\gamma [(1-\beta)\lambda_1 + \beta\lambda_2 + 1]} ,
\end{eqnarray}
where $\Gamma$ is the cluster decay time constant and
$\gamma$ is the intra-cluster decay time constant.
\item
The amplitudes $\{h_{k,c}\}$ are i.i.d Nakagami with $m$-factors
randomly generated via i.i.d $m\sim\mc{N}(m_0,\hat{m}^2_0)$.
\item
The phases $\{\phi_{k,c}\}$ are i.i.d uniform on $[0,2\pi)$.
\item
The number of clusters, $C$, is Poisson distributed with mean
$\bar{C}$, i.e., $p(C) \!=\! (\bar{C})^{C} \exp(-\bar{C}) / (C!)$.
\item
The number of components per cluster, $K$, is set large enough 
to yield a desired modeling accuracy.  
\end{itemize}
Beyond the above specifications, we assume the following.
\begin{itemize}
\item
The parameters
$\{\Lambda,\lambda_1,\lambda_2,\beta,\Gamma,\gamma,m_0,\hat{m}_0,\bar{C}\}$
are set according to the 802.15.4a ``outdoor NLOS'' scenario 
\cite{Molisch:802.15.4a}.
\item
$K\!=\!100$ components per cluster are used.
\item
The pulses $\gt(\tau)$ and $\gr(\tau)$ are square-root raised cosine 
(SRRC) designs with parameter $0.5$.
\item
The system bandwidth equals $T^{-1}=256$ MHz.
\item
The number of taps (and CP length) was set at $L=256$
(implying a maximal delay spread of $1\,\nu$sec) in order to capture all 
significant energy in $h(\tau)$. 
\item
The initial delay was generated via $T_0\!=\! \Lpre T+ \tilde{T}_0$,
where $\Lpre\!=\!20$ and where $\tilde{T}_0$ is exponentially 
distributed with mean $T$,
i.e., $p(\tilde{T}_0) \!=\! \Lambda_0 \exp(-\Lambda_0 \tilde{T}_0)$ for
$\Lambda_0 \!=\! 1/T$.
Here, $\Lpre$ was chosen so that $\{x_j\}_{j=0}^{\Lpre}$ captures
the ``pre-cursor'' energy contributed by the pulse shape, while
$\Lambda_0$ models a positive synchronization uncertainty.
%$\tau_{\min}\!=\!\Lpre T$.
\end{itemize}

We now show results from an experiment conducted 
using $U=10000$ realizations of the tap vector $\vec{x}$.
In \figref{hist}, we show histograms of $\real(x_j)$ for lags 
$j\!\in\!\{5,23,128,230\}$.
There it can be seen that the empirical distribution of $\real(x_j)$ 
changes significantly with lag $j$: 
for pre-cursor lags $j\!<\!\Lpre$, it is approximately Gaussian;
for near-cursor lags $j\!\approx\! \Lpre$, it is approximately
Laplacian; and, for post-cursor lags $j\!\gg\! \Lpre$, it is extremely heavy-tailed.
In \figref{realization}, we show a typical realization of $\vec{x}$ 
and notice clustering among the big taps. 
For comparison, we also plot an empirical estimate of the power-delay profile (PDP) 
$\vec{\rho}\defn[\rho_0,\dots,\rho_{L-1}]\tran$ in \figref{realization}, 
where $\rho_j \!\defn\! \E\{|x_j|^2\}$.

Next, we fit the GM2 parameters $\{\lambda_j,\nu^0_j,\nu^1_j\}_{j=0}^{L-1}$ 
from the realizations $\{\vec{x}_u\}_{u=1}^U$ using the EM algorithm 
\cite[p.~435]{Bishop:Book:07}, which iterates the steps 
\eqref{EM_post}-\eqref{EM_lam} until convergence: 
\begin{eqnarray}
  \omega_{j,u}
  &=& \textstyle \frac{\lambda_j \mc{CN}(x_{j,u}; 0,\nu_j^1)}
        {(1-\lambda_j) \mc{CN}(x_{j,u}; 0,\nu_j^0) 
        + \lambda_j \mc{CN}(x_{j,u}; 0,\nu_j^1)} ~\forall j,u
					\label{eq:EM_post}\\
  \nu^1_j &=& \textstyle \sum_{u=1}^U \omega_{j,u} |x_{j,u}|^2/
        \sum_{u=1}^U \omega_{j,u} ~\forall j\\
  \nu^0_j &=& \textstyle \sum_{u=1}^U (1-\omega_{j,u}) |x_{j,u}|^2/
        \sum_{u=1}^U (1-\omega_{j,u}) ~\forall j \quad\\
  \lambda_j &=& \textstyle \frac{1}{U} \sum_{u=1}^U \omega_{j,u} ~\forall j.	
					\label{eq:EM_lam}
\end{eqnarray}
Above, $\omega_{j,u}$ is the posterior on the state $d_{j,u}$ of tap $x_{j,u}$, 
i.e., 
$\omega_{j,u}=\Pr\{d_{j,u}\!=\!1\giv x_{j,u}; \lambda_j,\nu^0_j,\nu^1_j\}$.
The EM-estimated big-variance profile 
$\vec{\nu}^1\defn[\nu^1_0,\dots,\nu^1_{L-1}]\tran$ and 
small-variance profile $\vec{\nu}^0$ are shown in \figref{realization}, 
while the sparsity profile $\vec{\lambda}\defn[\lambda_0,\dots,\lambda_{L-1}]\tran$ 
is shown in \figref{gm2}.
Not surprisingly, the best-fit GM2 parameters also change significantly with lag $j$.
In particular, as $j$ becomes larger, the variance ratio $\nu^1_j/\nu^0_j$ 
increases while the big-tap-probability $\lambda_j$ decreases, corresponding to 
an increase in sparsity.
Meanwhile, there exists a peak in $\lambda_j$ near $j\!=\!\Lpre$ that results
from synchronization.

Next, we empirically estimated the switching probabilities 
$\vec{p}^{01}\defn[p^{01}_0,\dots,p^{01}_{L-1}]\tran$ and $\vec{p}^{10}$
using maximum a-posteriori (MAP) state estimates, i.e., 
$\hat{d}_{j,u} = \lfloor \omega_{j,u} + 0.5\rfloor$.
%To do this, we first detected the hidden state vector 
%$\vec{d}\defn [d_0,\dots,d_{L-1}]\tran$ underlying each realization of $\vec{x}$
%by comparing each element $x_j$ to the maximum a-posteriori (MAP) threshold 
%\cite{Poor:Book:94}
%%(i.e., the value $x$ such that $\Pr\{d_j\!=\!1\giv x_j\!=\!x\}\!=\!0.5$).
%(shown in \figref{realization}).
In particular,
\begin{eqnarray}
  p_j^{01} 
  &=& \textstyle \sum_{u=1}^U 1_{\{\hat{d}_{j+1,u}=0 \,\&\, \hat{d}_{j,u}=1\}} /
	\sum_{u=1}^U 1_{\{\hat{d}_{j,u}=1\}} \\
  p_j^{10} 
  &=& \textstyle \sum_{u=1}^U 1_{\{\hat{d}_{j+1,u}=1 \,\&\, \hat{d}_{j,u}=0\}} /
	\sum_{u=1}^U 1_{\{\hat{d}_{j,u}=0\}} ,
\end{eqnarray}
where $1_{\{A\}}$ denotes the indicator function for event $A$.
%From these detected states, estimates of $\vec{p}^{01}$ and $\vec{p}^{10}$ 
%were then computed by counting the number of switches (at each fixed lag $j$) 
%over the $10000$ realizations.
From the plots in \figref{gm2}, we see that the estimated switching 
probabilities are lag-dependent as well.

Finally, using the MAP state estimates $\{\hat{d}_{j,u}\}$, we empirically 
estimated the normalized conditional correlation 
%$$\textstyle \frac{ \E\{x_{j+1} x_j^*\giv d_{j+1}=1,d_j=1\} } 
%                  { \sqrt{\E\{|x_{j+1}|^2\giv d_{j+1}=1,d_j=1\} 
%       	                  \E\{|x_{j}|^2\giv d_{j+1}=1,d_j=1\}} } $$
$$\textstyle 
\frac{\sum_{u=1}^U 1_{\{\hat{d}_{j+1,u}=1,\hat{d}_{j,u}=1\}} 
	x_{j+1,u} x_{j,u}^* } 
     {\sqrt{\sum_{u=1}^U 1_{\{\hat{d}_{j+1,u}=1,\hat{d}_{j,u}=1\}} |x_{j+1,u}|^2
            \sum_{u=1}^U 1_{\{\hat{d}_{j+1,u}=1,\hat{d}_{j,u}=1\}} |x_{j,u}|^2}}
$$
and found that the magnitudes were $<\!0.1$, validating our assumption
of conditionally independent tap amplitudes.

In summary, we see that IEEE~802.15.4a channels do indeed
yield taps with the lag-dependent clustered sparsity 
described in \secref{GM2}. 
Moreover, we have shown how the GM2-HMM parameters can be estimated 
from realizations of $\vec{x}$. 
Next, we propose an efficient factor-graph based approach to joint 
channel-estimation and decoding (JCED) for BICM-OFDM using
the GM2-HMM prior proposed in \secref{GM2}.

\putFrag{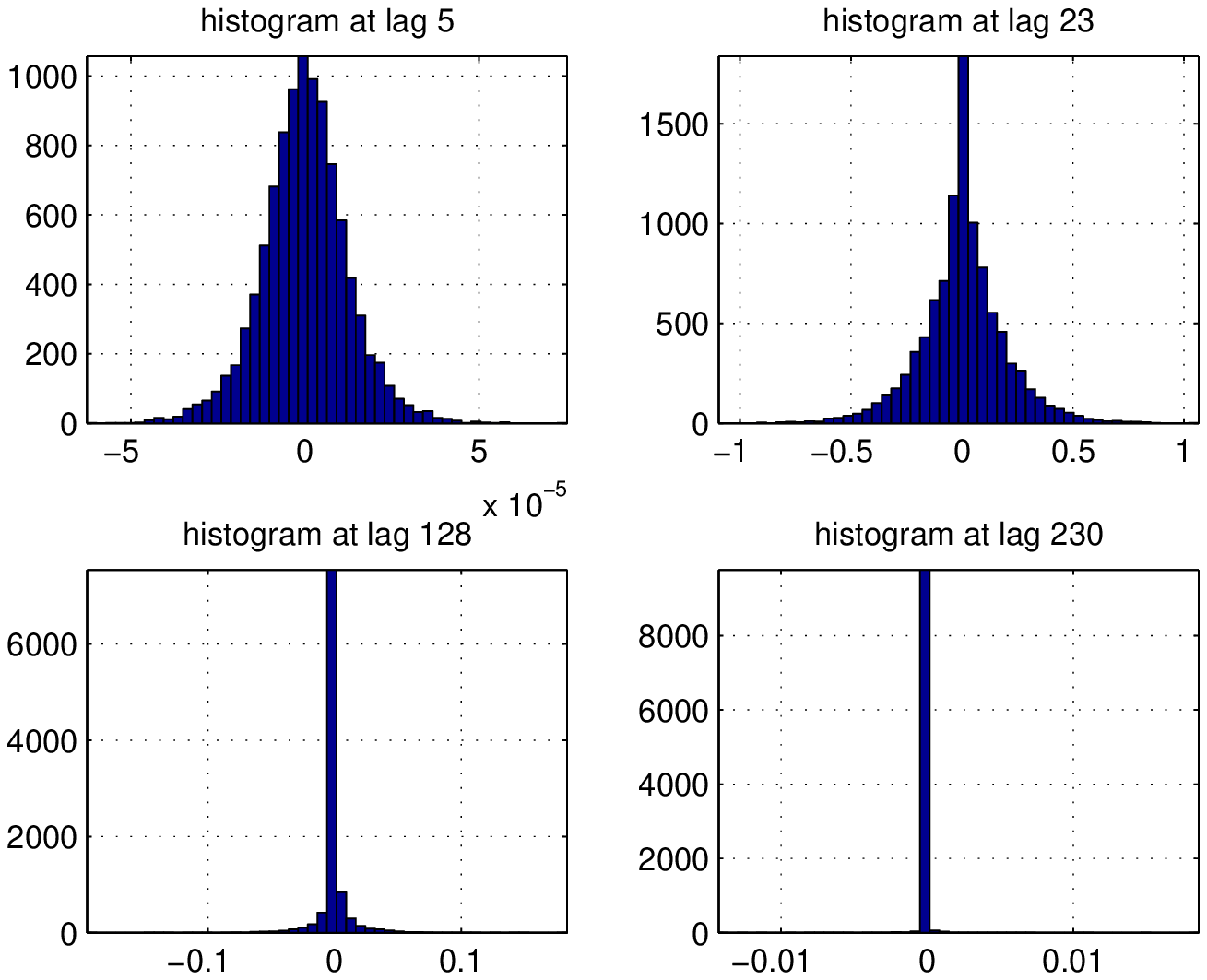}
	{Histograms of $\real(x_j)$ for lags $j\in\{5,23,128,230\}$, with 
 	 ``tight'' axes.  With synchronization delay $\Lpre\!=\!20$,
	 note that the histogram appears Gaussian for $j\!<\!\Lpre$,
	 Laplacian for $j\!\approx\! \Lpre$, and very sparse for $j\!\gg\! \Lpre$.}
	{\figsize}
	{}

\putFrag{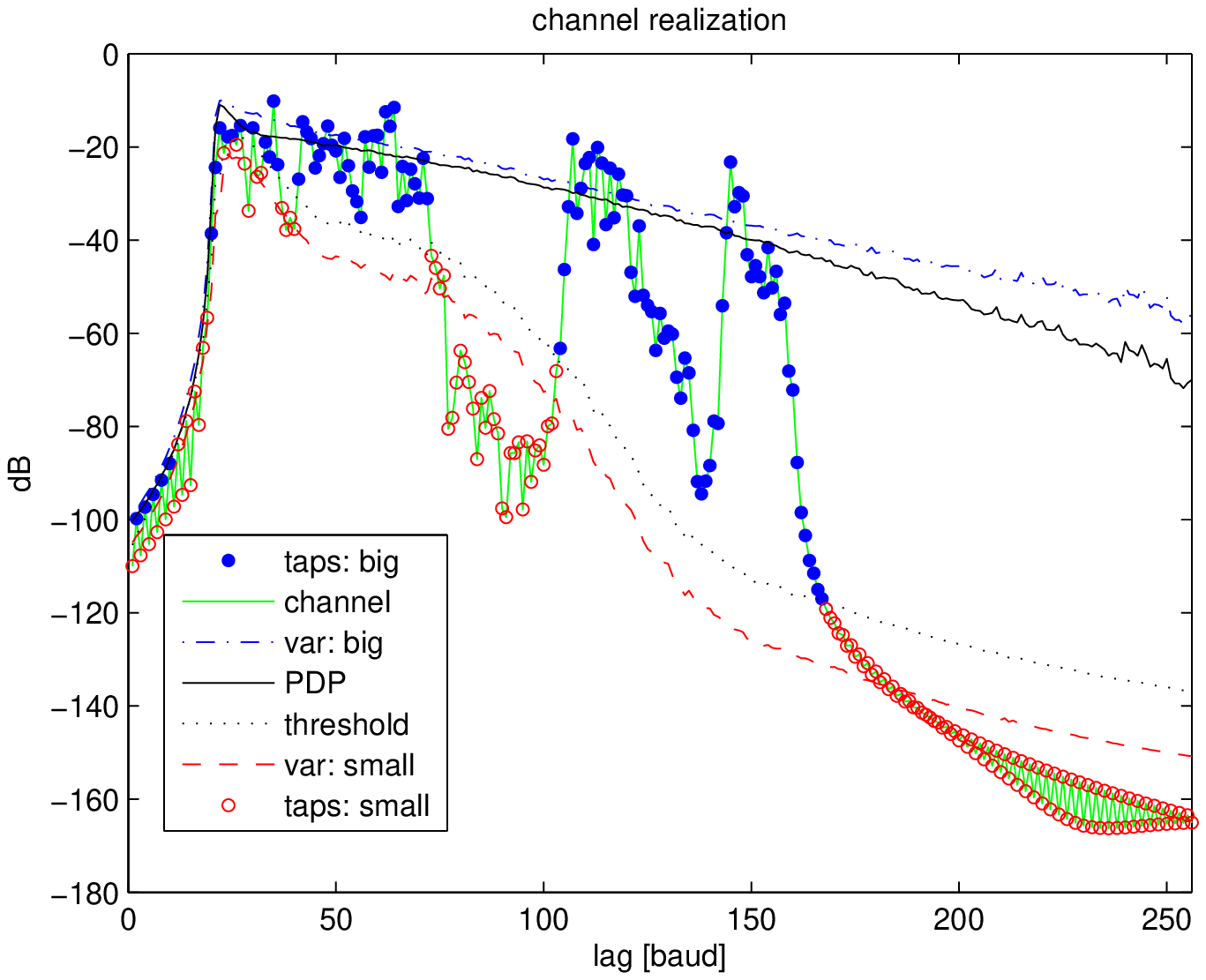}
	{A sample realization of channel taps $\{x_j\}$ generated from the 
	 IEEE~802.15.4a model with SRRC pulse shaping.  
	 Also shown is the empirically estimated PDP, best fits of the GM2 
	 parameters $\{\nu^0_j,\nu^1_j\}$, and the MAP threshold for detecting 
	 the hidden state $d_j$ given the tap value $x_j$.}
	{\figsize}
	{\newcommand{\sz}{0.7}
	 \psfrag{channel realization}[b][b][1.0]{}
	 \psfrag{lag [baud]}[][][\sz]{\sf lag $j$ [baud]} }

\putFrag{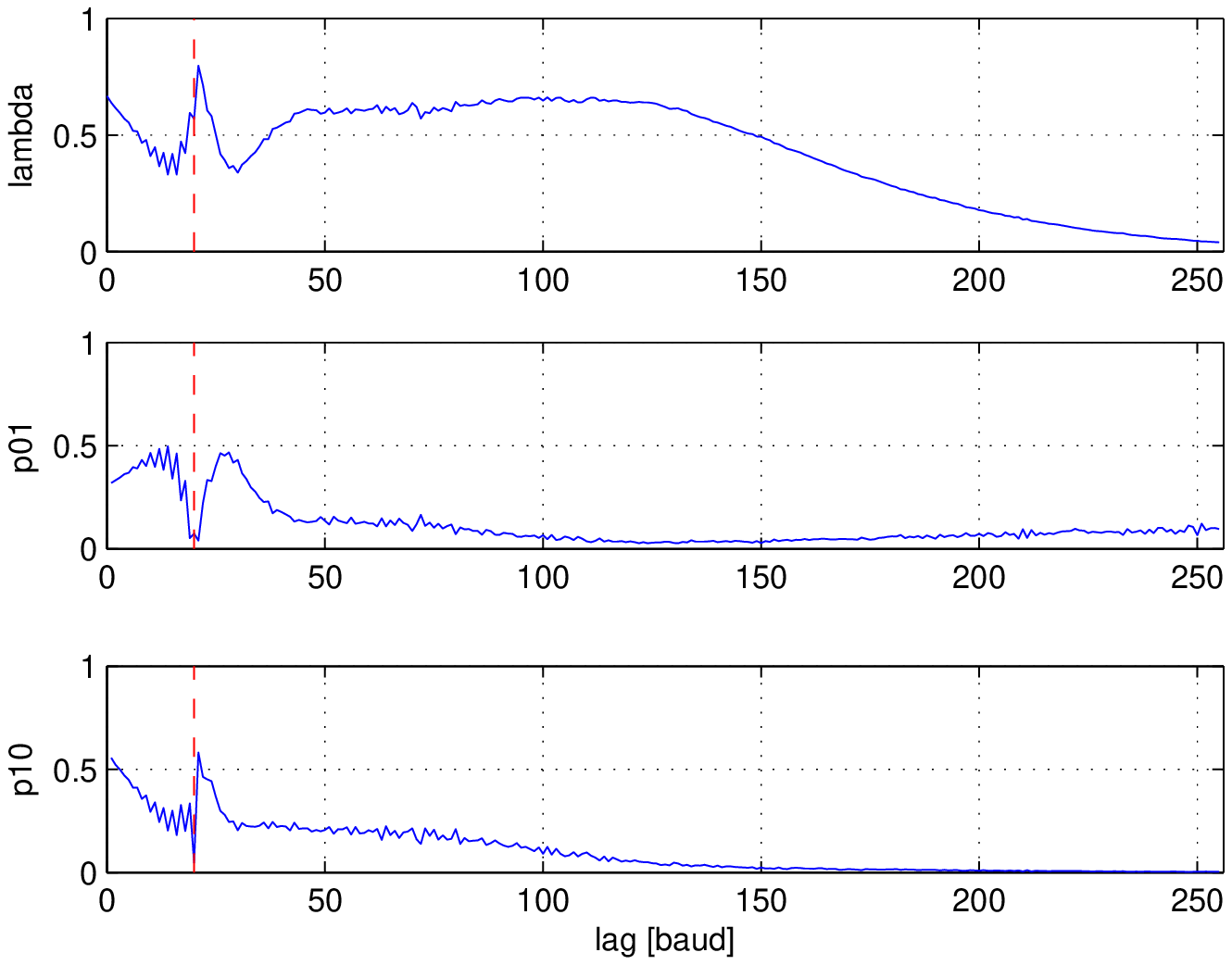}	
	{Empirically estimated statistics on the tap-states $\{d_j\}$.
	Top: $\lambda_j\defn \Pr\{d_j\!=\!1\}$,
	middle: $p^{01}_j \defn \Pr\{d_{j+1}\!=\!0\giv d_j\!=\!1\}$,
	bottom: $p^{10}_j \defn \Pr\{d_{j+1}\!=\!1\giv d_j\!=\!0\}$.
	The red dashed line shows the synchronization reference, $j=\Lpre=20$.}
	{\figsize}
	{\newcommand{\sz}{0.7}
	 \psfrag{lambda}[b][b][\sz]{$\lambda_j$}
	 \psfrag{p01}[][][\sz]{$p^{01}_j$}
	 \psfrag{p10}[][][\sz]{$p^{10}_j$}  
	 \psfrag{lag [baud]}[][][\sz]{\sf lag $j$ [baud]} }

%%%%%%%%%%%%%%%%%%%%%%%%%%%%%%%%%%%%%%%%%%%%%%%%%%%%%%%%%%%%%%%%%%%%%%%%%%%%%
\section{Joint Channel Estimation and Decoding}	\label{sec:jced}

\putFrag{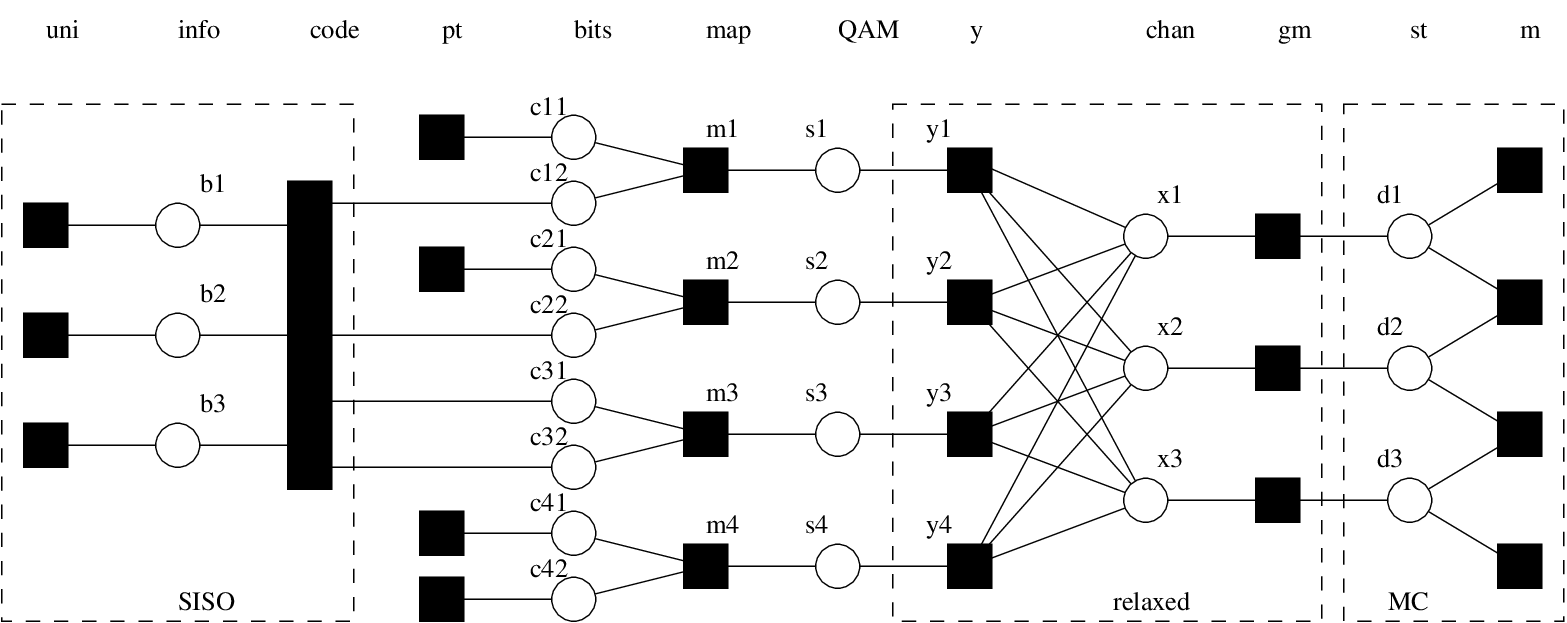}
	{Factor graph of the JCED problem for a 
	 toy example with $\Mi=3$ information bits, $\Np=1$ pilot subcarrier
	 (at subcarrier index $i=3$), $\Mt=2$ training bits, $M=2$ bits per 
	 QAM symbol, $N=4$ OFDM subcarriers, and channel impulse response 
	 length $L=3$.}
 	{3.5}
	{\newcommand{\vs}{0.8}
	 \newcommand{\cs}{0.7}
	 \newcommand{\ts}{0.50}
	 \psfrag{SISO}[B][Bl][\ts]{\sf SISO decoding}
	 \psfrag{relaxed}[B][Bl][\ts]{\sf GAMP}
	 \psfrag{MC}[B][Bl][\ts]{\sf MC}
	 \psfrag{b1}[Bl][Bl][\vs]{$b_1$}
	 \psfrag{b2}[Bl][Bl][\vs]{$b_2$}
	 \psfrag{b3}[Bl][Bl][\vs]{$b_3$}
	 \psfrag{b4}[Bl][Bl][\vs]{$b_4$}
	 \psfrag{c11}[B][Bl][\cs]{$c_{0,1}$}
	 \psfrag{c21}[B][Bl][\cs]{$c_{1,1}$}
	 \psfrag{c31}[B][Bl][\cs]{$c_{2,1}$}
	 \psfrag{c41}[B][Bl][\cs]{$c_{3,1}$}
	 \psfrag{c12}[B][Bl][\cs]{$c_{0,2}$}
	 \psfrag{c22}[B][Bl][\cs]{$c_{1,2}$}
	 \psfrag{c32}[B][Bl][\cs]{$c_{2,2}$}
	 \psfrag{c42}[B][Bl][\cs]{$c_{3,2}$}
	 \psfrag{m1}[b][Bl][\vs]{$\mc{M}_0$}
	 \psfrag{m2}[b][Bl][\vs]{$\mc{M}_1$}
	 \psfrag{m3}[b][Bl][\vs]{$\mc{M}_2$}
	 \psfrag{m4}[b][Bl][\vs]{$\mc{M}_3$}
	 \psfrag{s1}[B][Bl][\vs]{$s_0$}
	 \psfrag{s2}[B][Bl][\vs]{$s_1$}
	 \psfrag{s3}[B][Bl][\vs]{$s_2$}
	 \psfrag{s4}[B][Bl][\vs]{$s_3$}
	 \psfrag{y1}[B][Bl][\vs]{$y_0$}
	 \psfrag{y2}[B][Bl][\vs]{$y_1$}
	 \psfrag{y3}[B][Bl][\vs]{$y_2$}
	 \psfrag{y4}[B][Bl][\vs]{$y_3$}
	 \psfrag{x1}[Bl][Bl][\vs]{$x_1$}
	 \psfrag{x2}[Bl][Bl][\vs]{$x_2$}
	 \psfrag{x3}[Bl][Bl][\vs]{$x_3$}
	 \psfrag{d1}[B][Bl][\vs]{$d_1$}
	 \psfrag{d2}[B][Bl][\vs]{$d_2$}
	 \psfrag{d3}[B][Bl][\vs]{$d_3$}
	 \psfrag{uni}[][Bl][\ts]{\sf
	 	\begin{tabular}{@{}c@{}}uniform\\[-1mm]prior\end{tabular}}
	 \psfrag{info}[][Bl][\ts]{\sf
	 	\begin{tabular}{@{}c@{}}info\\[-1mm]bits\end{tabular}}
	 \psfrag{code}[][Bl][\ts]{\sf
	 	\begin{tabular}{@{}c@{}}code \&\\[-1mm]interlv\end{tabular}}
	 \psfrag{pt}[][Bl][\ts]{\sf
	 	\begin{tabular}{@{}c@{}}pilots \&\\[-1mm]training\end{tabular}}
	 \psfrag{bits}[][Bl][\ts]{\sf
	 	\begin{tabular}{@{}c@{}}coded\\[-1mm]bits\end{tabular}}
	 \psfrag{map}[][Bl][\ts]{\sf
	 	\begin{tabular}{@{}c@{}}symbol\\[-1mm]mapping\end{tabular}}
	 \psfrag{QAM}[][Bl][\ts]{\sf
	 	\begin{tabular}{@{}c@{}}QAM\\[-1mm]symbs\end{tabular}}
	 \psfrag{y}[][Bl][\ts]{\sf
	 	\begin{tabular}{@{}c@{}}OFDM\\[-1mm]observ\end{tabular}}
	 \psfrag{chan}[][Bl][\ts]{\sf
	 	\begin{tabular}{@{}c@{}}channel\\[-1mm]taps\end{tabular}}
	 \psfrag{gm}[][Bl][\ts]{\sf
	 	\begin{tabular}{@{}c@{}}sparse\\[-1mm]prior\end{tabular}}
	 \psfrag{st}[][Bl][\ts]{\sf
	 	\begin{tabular}{@{}c@{}}tap\\[-1mm]states\end{tabular}}
	 \psfrag{m}[][Bl][\ts]{\sf
	 	\begin{tabular}{@{}c@{}}cluster\\[-1mm]prior\end{tabular}}
	 }

Our goal is to infer the information bits $\vec{b}$
from the OFDM observations $\vec{y}$ 
and the pilot/training bits $\vec{c}\pt$, 
without knowing the channel state $\vec{x}$. 
In particular, we aim to maximize the posterior pmf 
$p(b_m\giv\vec{y},\vec{c}\pt)$ of each info bit.
To exploit prior knowledge that $\vec{x}$ is clustered-sparse, we 
employ the GM2-HMM prior described in \secref{GM2}.
As a result, the info-bit posterior can be decomposed into
the following product of factors:
\begin{eqnarray}
  \lefteqn{ 
  p(b_m\giv \vec{y},\vec{c}\pt) 
  \,=\, \sum_{\vec{b}_{-m}} 
  	p(\vec{b}\giv \vec{y},\vec{c}\pt) 
  \,\propto\, 
  	\sum_{\vec{b}_{-m}} 
  	p(\vec{y}\giv \vec{b},\vec{c}\pt) p(\vec{b}) 
  } \quad \label{eq:propto} \\
  &=& \int_{\vec{x}} 
	\sum_{\vec{s},\vec{d},\vec{c},\vec{b}_{-m}} \hspace{-4mm}
  	p(\vec{y}\giv \vec{s}, \vec{x}) 
	p(\vec{x}\giv \vec{d}) p(\vec{d})
	p(\vec{s}\giv \vec{c})
	p(\vec{c}\giv \vec{b},\vec{c}\pt) p(\vec{b}) 	\nonumber\\
  &=& \int_{\vec{x}} \sum_{\vec{d}} 
	\prod_{j=0}^{L-1} p(x_j\giv d_j) p(d_j\giv d_{j-1})
	\sum_{\vec{s}} 
  	\prod_{i=0}^{N-1} p(y_i\giv s_i, \vec{x}) 
	\nonumber\\&&\mbox{}\times
  	\sum_{\vec{c}}
	%\prod_{i'=0}^{N-1} p(s_{i'}\giv\vec{c}_{i'})
	p(s_{i}\giv\vec{c}_{i})
  	\sum_{\vec{b}_{-m}} 
	p(\vec{c}\giv\vec{b},\vec{c}\pt) 
	\prod_{m=1}^{\Mi} p(b_{m}),\quad	\label{eq:factored}
\end{eqnarray}
where 
$\vec{b}_{-m}\!\defn\![b_1,\dots,b_{m-1},b_{m+1},\dots,b_{\Mi}]\tran$.
This factorization is illustrated by the \emph{factor 
graph} in \figref{factor_graph_noncoh_clust}, where the round nodes represent
random variables and the square nodes represent the factors of the 
posterior exposed in \eqref{factored}.

\subsection{Background on belief propagation}	\label{sec:bp}

Although exact evaluation of the posteriors $\{p(b_m\giv\vec{y},\vec{c}\pt)\}$
is computationally impractical for the problem sizes of interest,
these posteriors can be approximately evaluated using 
\emph{belief propagation} (BP) \cite{Frey:ANIPS:98} on the factor graph in 
\figref{factor_graph_noncoh_clust}.
In textbook BP, beliefs take the form of pdfs/pmfs that are propagated 
among nodes of the factor graph via the \emph{sum/product algorithm} (SPA)
\cite{Pearl:Book:88,Loeliger:Proc:07,Kschischang:TIT:01}:
\begin{enumerate}
 \item
 Say the factor node $f$ is connected to the variable nodes 
 $\{v_a\}_{a=1}^A$. The belief passed from $f$ to $v_b$ is 
 $p_{f\rightarrow v_b}(v_b) 
 \propto \int_{\{v_a\}_{a\neq b}} f(v_1,\dots,v_A) 
 \prod_{a\neq b} p_{v_a \rightarrow f}(v_a)$,
 given the beliefs $\{p_{v_a\rightarrow f}(\cdot)\}_{a\neq b}$ recently 
 passed to $f$.
 \item
 Say the variable node $v$ is connected to the factor nodes $\{f_1,\dots,f_B\}$.
 The belief passed from $v$ to $f_a$ is 
 $p_{v\rightarrow f_a}(v)
 \propto \prod_{b\neq a} p_{f_b\rightarrow v}(v)$,
 given the beliefs $\{p_{f_b\rightarrow v}(\cdot)\}_{b\neq a}$ recently 
 passed to $v$.
 \item
 Say the variable node $v$ is connected to the factor nodes $\{f_1,\dots,f_B\}$.
 The posterior on $v$ is the product of all recently arriving beliefs,
 i.e., $p(v) \propto \prod_{b=1}^B p_{f_b\rightarrow v}(v)$.
\end{enumerate}
%\Figref{bp} helps to illustrate the first two rules.

When the factor graph contains no loops, SPA-BP yields exact posteriors after
two rounds of message passing (i.e., forward and backward). 
But, in the presence of loops, convergence to the exact posteriors is not guaranteed \cite{Cooper:AI:90}.
That said, there exist many problems to which loopy BP \cite{Frey:ANIPS:98} 
has been successfully applied, including
%turbo decoding \cite{McEliece:JSAC:98},
%multiuser detection \cite{Boutros:TIT:02},
inference on Markov random fields \cite{Freeman:IJCV:00},
LDPC decoding \cite{MacKay:Book:03},
and compressed sensing 
\cite{Baron:TSP:10,Donoho:PNAS:09,Bayati:10,Rangan:10b,Schniter:CISS:10,Rangan:10v2}. 
Our work not only leverages these past successes, but unites them.

%\putFrag{bp}
%	{Examples of belief propagation among nodes of a factor graph.}
%	{3.25}
%	{\psfrag{v1}[r][Bl][1.0]{$v_1$}
%	 \psfrag{v2}[r][Bl][1.0]{$v_2$}
%	 \psfrag{v3}[l][Bl][1.0]{$v_3$}
%	 \psfrag{f}[][Bl][1.0]{$f$}
%	 \psfrag{pv1>f}[Bl][Bl][1.0]{$p_{v_1\rightarrow f}$}
%	 \psfrag{pv2>f}[tl][Bl][1.0]{$p_{v_2\rightarrow f}$}
%	 \psfrag{pf>v3}[B][Bl][1.0]{$p_{f\rightarrow v_3}$}
%	 \psfrag{f1}[r][Bl][1.0]{$f_1$}
%	 \psfrag{f2}[r][Bl][1.0]{$f_2$}
%	 \psfrag{f3}[l][Bl][1.0]{$f_3$}
%	 \psfrag{v}[t][Bl][1.0]{$v$}
%	 \psfrag{pf1>v}[Bl][Bl][1.0]{$p_{f_1\rightarrow v}$}
%	 \psfrag{pf2>v}[tl][Bl][1.0]{$p_{f_2\rightarrow v}$}
%	 \psfrag{pv>f3}[B][Bl][1.0]{$p_{v \rightarrow f_3}$}
%	}

\subsection{Background on GAMP}			\label{sec:gamp}

An important sub-problem within our larger bit-inference problem is the 
estimation of a vector of independent possibly-non-Gaussian variables 
$\vec{x}$ that are linearly mixed via $\vec{\Phi}\in\Complex^{N\times L}$ 
to form $\vec{z}\!=\!\vec{\Phi x}\!=\![z_0,\dots,z_{N-1}]\tran$,
and subsequently observed as noisy measurements $\vec{y}$ through the 
possibly non-Gaussian pdfs $\{p_{Y_i|Z_i}(.\giv.)\}_{i=0}^{N-1}$.
In our case, \eqref{pxj} specifies a GM2 prior on $x_j$ 
and \eqref{yi}---given the finite-alphabet uncertainty in $s_i$---yields the 
non-Gaussian measurement pdf $p_{Y_i|Z_i}$.
This ``linear mixing'' sub-problem is described by the factor graph shown within the 
middle dashed box in \figref{factor_graph_noncoh_clust},
where each node ``$y_i$'' represents the measurement pdf
$p_{Y_i|Z_i}$ and the node rightward of each node ``$x_j$'' 
represents the GM2 prior on $x_j$.

Building on recent work on multiuser detection by Guo and Wang 
\cite{Guo:ISIT:07}, as well as recent work on message passing algorithms
for compressed sensing by Donoho, Maleki, Montanari, and Bayati
\cite{Donoho:PNAS:09,Bayati:10}, Rangan proposed a so-called 
\emph{generalized approximate message passing} (GAMP) scheme 
that, for the sub-problem described above, 
admits rigorous analysis\footnote{
  Since it is difficult to give a concise yet accurate account of GAMP's 
  technical properties, we refer the interested reader to \cite{Rangan:10b}.}
as $N,L\!\rightarrow\!\infty$ \cite{Rangan:10b}.
The main ideas behind GAMP are the following.
First, although the beliefs flowing leftward from the nodes $\{x_j\}$ 
are clearly non-Gaussian, the corresponding belief about 
$z_i = \sum_{j=0}^{L-1}\Phi_{ij} x_j$ can be accurately approximated as 
Gaussian, when $L$ is large, using the central limit theorem.
Moreover, to calculate the parameters of this distribution (i.e., its
mean and variance), only the mean and variance of each $x_j$ are needed.
Thus, it suffices to pass only means and variances leftward from each
$x_j$ node.
It is similarly desirable to pass only means and variances rightward
from each measurement node.
Although the exact rightward flowing beliefs would be non-Gaussian (due to
the non-Gaussian assumption on the measurement channels $p_{Y_i|Z_i}$),
GAMP approximates them as Gaussian using a 2nd-order Taylor series,
and passes only the resulting means and variances.
A further simplification employed by GAMP is to approximate
the \emph{differences} among the outgoing means/variances of each left node,
and the incoming means/variances of each right node, using Taylor series.
The GAMP algorithm\footnote{
  To be precise, the GAMP algorithm in \tabref{gamp} is an extension of 
  that proposed in \cite{Rangan:10b}.
  \tabref{gamp} handles circular \emph{complex-valued} distributions
  and \emph{non-identically} distributed signals and measurements.} 
is summarized in \tabref{gamp}.

\putTable{gamp}{The GAMP Algorithm}{
\begin{equation*}
\begin{array}{|lrcl@{}r|}\hline
  \multicolumn{2}{|l}{\textsf{definitions:}}&&&\\[-1mm]
  &p_{Z_i|Y_i}(z|y;\hat{z},\nu^z)
   &=& \frac{p_{Y_i|Z_i}(y|z) \,\mc{CN}(z;\hat{z},\nu^z)}
	{\int_{z'} p_{Y_i|Z_i}(y|z') \,\mc{CN}(z';\hat{z},\nu^z)} &\text{(D1)}\\
  &g\out(y,\hat{z},\nu^z)
   &=& \frac{1}{\nu^z} \left(\E_{Z_i|Y_i}\{ z|y;\hat{z},\nu^z\}-\hat{z}\right) &\text{(D2)}\\
  &g\out'(y,\hat{z},\nu^z)
   &=& \frac{1}{\nu^z}\left(\frac{\var_{Z_i|Y_i}\{z|y;\hat{z},\nu^z\}}{\nu^z}-1\right)&\text{(D3)}\\
  &p_{X_j}\!(x;\hat{r},\nu^r)
   &=& \frac{p_{X_j}\!(x) \,\mc{CN}(x;\hat{r},\nu^r)}
        {\int_{x'}p_{X_j}\!(x') \,\mc{CN}(x';\hat{r},\nu^r)}&\text{(D4)}\\
  &g\inp(\hat{r},\nu^r)
   &=& \int_x x\, p_{X_j}\!(x;\hat{r},\nu^r) &\text{(D5)}\\
  &g\inp'(\hat{r},\nu^r)
   &=& \frac{1}{\nu^r}\int_x |x-g\inp(\hat{r},\nu^r)|^2\, p_{X_j}\!(x;\hat{r},\nu^r)\quad &\text{(D6)}\\
  \multicolumn{2}{|l}{\textsf{initialize:}}&&&\\
  &\forall j: 
   \hat{x}_{j}(1) &=& \int_{x} x\, p_{X_j}(x) & \text{(I1)}\\
  &\forall j:
   \nu^x_j(1) &=& \int_{x} |x-\hat{x}_j(1)|^2  p_{X_j}(x) & \text{(I2)}\\
  &\forall i: 
   \hat{u}_{i}(0) &=& 0 & \text{(I3)}\\
  \multicolumn{2}{|l}{\textsf{for $n=1,2,3,\dots$}}&&&\\
  &\forall i:
   \hat{z}_i(n)
   &=& \textstyle \sum_{j=0}^{L-1} \Phi_{ij} \hat{x}_{j}(n) & \text{(R1)}\\
  &\forall i:
   \nu^z_i(n)
   &=& \textstyle \sum_{j=0}^{L-1} |\Phi_{ij}|^2 \nu^x_{j}(n) & \text{(R2)}\\
  &\forall i:
   \hat{p}_i(n)
   &=& \hat{z}_{i}(n) - \nu^z_i(n) \,\hat{u}_i(n-1)& \text{(R3)}\\
  &\forall i:
   \hat{u}_i(n)
   &=& g\out(y_i,\hat{p}_i(n),\nu^z_i(n)) & \text{(R4)}\\
  &\forall i:
   \nu^u_i(n)
   &=& -g'\out(y_i,\hat{p}_i(n),\nu^z_i(n)) & \text{(R5)}\\
  &\forall j:
   \nu^r_j(n)
   &=& \textstyle \big(\sum_{i=0}^{N-1} |\Phi_{ij}|^2 \nu^u_i(n) 
	\big)^{-1} & \text{(R6)}\\
  &\forall j:
   \hat{r}_j(n)
   &=& \textstyle \hat{x}_j(n)+ \nu^r_j(n) \sum_{i=0}^{N-1} \Phi_{ij}^*
	\hat{u}_{i}(n)  & \text{(R7)}\\
  &\forall j:
   \nu^x_j(n\!+\!1)
   &=& \nu^r_j(n) g'\inp(\hat{r}_j(n),\nu^r_j(n)) & \text{(R8)}\\
  &\forall j:
   \hat{x}_{j}(n\!+\!1)
   &=& g\inp(\hat{r}_j(n),\nu^r_j(n)) & \text{(R9)}\\
  \multicolumn{2}{|l}{\textsf{end}}&&&\\\hline
\end{array}
\end{equation*}
}

\subsection{Joint estimation and decoding using GAMP}		\label{sec:jced_gamp}

We now detail our application of GAMP to joint channel-estimation and decoding 
(JCED) under the GM2-HMM tap prior,
frequently referring to the factor graph in \figref{factor_graph_noncoh_clust}.

Because our factor graph is loopy, there exists considerable 
freedom in the message passing schedule.
Roughly speaking, we choose to pass messages from the left to the right 
of \figref{factor_graph_noncoh_clust} and back again,
several times, stopping as soon as the messages converge. 
Each of these full cycles of message passing will be referred to as a 
``turbo iteration.'' 
However, during a single turbo iteration, there may be multiple iterations
of message passing \emph{between} the GAMP and MC sub-graphs, 
which will be referred to as ``equalizer'' iterations.
Furthermore, during a single equalizer iteration, there may be
multiple iterations of message passing \emph{within} the GAMP sub-graph,
while there is at most one forward-backward iteration \emph{within} 
the MC sub-graph. 
Finally, the SISO decoding block may itself be implemented using message
passing, in which case it may also use several internal iterations.
The message passing details are discussed below.

At the start of the first turbo iteration, 
there is total uncertainty about the information bits,
so that $\Pr\{b_m\!=\!1\}\!=\!\frac{1}{2}~\forall m$. 
Thus, the initial bit beliefs flowing rightward
out of the coding/interleaving block are uniformly distributed. 
%(i.e., $p_{c_{i,m}\rightarrow \delta_i}(c)=\frac{1}{2}~\forall c$ for
%all $(i,m)$ corresponding to information bits).
Meanwhile, the pilot/training bits are known with certainty. 
%so that
%$p_{c_{i,m}\rightarrow \delta_i}(c)=1$ for $c=c_{i,m}$ when $(i,m)$ correspond
%to pilot/training bits.

Coded-bit beliefs are then propagated rightward into the symbol mapping nodes. 
Since the symbol mapping is deterministic, the corresponding pdf factors 
take the form $p(s\of{k}\giv\vec{c}\of{l}) = \delta_{k-l}$. 
The SPA dictates that the message passed rightward from symbol mapping 
node ``$\mc{M}_i$'' takes the form
\begin{eqnarray}
  p_{\mc{M}_i\rightarrow s_i}(s\of{k})
  &\propto& \sum_{\vec{c}\in\{0,1\}^M} p(s\of{k}|\vec{c}) \prod_{m=1}^M 
  	p_{c_{i,m}\rightarrow \mc{M}_i}(c_m) \quad \\
  &=& \prod_{m=1}^M p_{c_{i,m}\rightarrow \mc{M}_i}(c_m\of{k}) ,
\end{eqnarray}
which is then copied forward as the message passed rightward from node $s_i$ 
(i.e., $p_{\mc{M}_i\rightarrow s_i}(s\of{k})=p_{s_i\rightarrow y_i}(s\of{k})$).

Recall, from \secref{gamp}, that the symbol-belief passed rightward into the
measurement node ``$y_i$'' determines the pdf $p_{Y_i|Z_i}$ used in GAMP.
Writing this symbol belief as 
$\vec{\beta}_i\defn[\beta_i\of{1},\dots,\beta_i\of{|\const|}]\tran$ for
$\beta_i\of{k}\defn p_{s_i\rightarrow y_i}(s\of{k})$,
equation \eqref{yi} implies the measurement pdf 
\begin{eqnarray}
  p_{Y_i|Z_i}(y|z)
  &=& \sum_{k=1}^{|\const|} \beta_i\of{k} \,\mc{CN}(y;s\of{k}z;\nu^w) .
  							\label{eq:pY|Z}
\end{eqnarray}
From \eqref{pY|Z}, it is shown in \appref{out} that the 
quantities in (D2)-(D3) of \tabref{gamp} become
\begin{eqnarray}
  g\out(y,\hat{z},\nu^z)
  &=& \frac{1}{\nu^z} \hat{e}_i(y,\hat{z},\nu^z) 		\label{eq:gout}\\
  g'\out(y,\hat{z},\nu^z)
  &=& \frac{1}{\nu^z}\Big(\frac{\nu^e_i(y,\hat{z},\nu^z)}{\nu^z}-1\Big) 		
  								\label{eq:g'out}
\end{eqnarray}
for
\begin{eqnarray}
  \xi_i\of{k}(y,\hat{z},\nu^z)
  &\defn& \frac{\beta_i\of{k}
  		\mc{CN}(y;s\of{k}\hat{z},|s\of{k}|^2\nu^z\!+\!\nu^w)}
           {\sum_{k'} \beta_i\of{k'}
	    	\mc{CN}(y;s\of{k'}\hat{z},|s\of{k'}|^2\nu^z\!+\!\nu^w)} 
	    							\quad\label{eq:xi}\\
  \zeta\of{k}(\nu^z)
  &\defn& \frac{|s\of{k}|^2\nu^z}{|s\of{k}|^2\nu^z+\nu^w} 	\label{eq:zeta}\\
  \hat{e}\of{k}(y,\hat{z},\nu^z)
  &\defn& \Big(\frac{y}{s\of{k}}-\hat{z}\Big) \zeta\of{k}(\nu^z) \\
  \hat{e}_i(y,\hat{z},\nu^z)
  &\defn& \sum_{k=1}^{|\const|} \xi_i\of{k}\!(y,\hat{z},\nu^z)
  	\,\hat{e}\of{k}(y,\hat{z},\nu^z) 			\label{eq:ei} \\
  \nu^{e}_i(y,\hat{z},\nu^z)
  &\defn& \sum_{k=1}^{|\const|} \xi_i\of{k}\!(y,\hat{z},\nu^z)
  	\,\bigg( |\hat{e}\of{k}(y,\hat{z},\nu^z) - \hat{e}_i(y,\hat{z},\nu^z)|^2
		+ \frac{\nu^w \zeta\of{k}(\nu^z)}{s\of{k}} \bigg)
								\label{eq:muei} 
\end{eqnarray}
where $\vec{\xi}_i\!\defn\![\xi_i\of{1},\dots,\xi_i\of{|\const|}]\tran$ characterizes the
posterior pmf on $s_i$ under the channel model $z_i\sim\mc{CN}(\hat{z},\nu^z)$.
Likewise, from \eqref{pxj}, it is shown in \appref{in} that the quantities (D5)-(D6) 
take the form 
\begin{eqnarray}
  g\inp(\hat{r},\nu^r)
  &=& \Big( \alpha_j \,\gamma^1_j + 
  	\big(1-\alpha_j\big) \,\gamma^0_j \Big)\hat{r} \label{eq:gin}\\
  g'\inp(\hat{r},\nu^r)
  &=& \alpha_j(1-\alpha_j)(\gamma^1_j-\gamma^0_j)^2 \,|\hat{r}|^2/\nu^r
%\nonumber\\&&\mbox{}
  	+ \alpha_j\gamma^1_j + (1-\alpha_j)\gamma^0_j ,
								\label{eq:g'in}
\end{eqnarray}
for 
$\gamma^0_j(\nu^r) \defn (1+\nu^r/\nu_j^0)^{-1}$ and	
$\gamma^1_j(\nu^r) \defn (1+\nu^r/\nu_j^1)^{-1}$ and
\begin{eqnarray}
  \alpha_j(\hat{r},\nu^r)
  &\defn& \frac{1}{1+\Bigg(
  	\underbrace{ \frac{\lambda_j}{1-\lambda_j} }_{ \displaystyle \mc{L}\pri_j }
  	\underbrace{ \frac{\mc{CN}(\hat{r};0,\nu_j^1+\nu^r)}
	        {\mc{CN}(\hat{r};0,\nu_j^0+\nu^r)} }_{ \displaystyle \mc{L}\ext_j(\hat{r},\nu^r) }
	\Bigg)^{-1}} .	
  							\label{eq:alfj}
\end{eqnarray}
Above,
$\mc{L}_j\pri$ is the apriori likelihood ratio $\frac{\Pr\{d_j=1\}}{\Pr\{d_j=0\}}$ 
on the hidden state,
$\mc{L}_j\ext(\hat{r},\nu^r)$ is GAMP's extrinsic likelihood ratio, and 
$\alpha_j(\hat{r},\nu^r)$ is the corresponding posterior probability that $d_j=1$.

Using \eqref{gout}-\eqref{alfj}, the GAMP algorithm in \tabref{gamp} is
iterated until it converges.\footnote{
  More precisely, GAMP is iterated until the mean-square tap-estimate 
  difference $\frac{1}{L}\sum_{j=0}^{L-1}|\hat{x}_j(n)-\hat{x}_j(n-1)|^2$ 
  falls below a threshold or a maximum number of GAMP iterations has elapsed.}
In doing so, GAMP generates (a close approximation to) both the 
conditional means $\hvec{x}$ and variances 
$\vec{\nu}^x\!\defn\![\nu_0^x,\dots,\nu_{L-1}^x]\tran$
given the observations $\vec{y}$, the soft symbol priors
$\vec{\beta}\!\defn\![\vec{\beta}_{0},\dots,\vec{\beta}_{L-1}]\tran$ 
and the sparsity prior $\vec{\lambda}$. 
Conveniently, GAMP also returns (close approximations to) both the
conditional means $\hvec{z}$ and variances $\vec{\nu}^z$ 
of the subchannel gains $\vec{z}$, as well as posteriors
$\vec{\xi}\!\defn\![\vec{\xi}_{0},\dots,\vec{\xi}_{L-1}]\tran$
on the symbols $\vec{s}$.

Before continuing, we discuss some GAMP details that are specific to our
OFDM-JCED application.
First, we notice that, to guarantee that the variance $\nu_i^u(n)$ in (R5) 
is positive, we must have $\nu^e_i\!<\!\nu_z$ in \eqref{g'out}.
Since this is not necessarily the case during the first few GAMP iterations,
we clip $\nu^e_i$ at the value $0.99\nu^z$, where $0.99$ was chosen heuristically.
Second, due to unit-modulus property of the DFT elements $\Phi_{ij}$,
step (R2) in \tabref{gamp} simplifies to $\nu^z_i(n) \!=\! \sum_j \nu_j^x(n)$ 
and (R6) simplifies to $\nu_j^r(n) \!=\! \big(\sum_i \nu_i^u(n)\big)^{-1}$.
With these simplifications, the complexity of GAMP is dominated by either
the matrix-vector products $\sum_j \Phi_{ij}\hat{x}_{j}(n)$ 
in (R1) and $\sum_i \Phi_{ij}^*\hat{u}_{i}(n)$ in (R7), which can be implemented
using a $N\log_2 N$-multiply FFT when $N$ is a power-of-two, 
or by the calculation of $\{\hat{e}_i,\nu_i^e\}_{i=0}^{N-1}$ in 
\eqref{ei}-\eqref{muei}, which requires $\mc{O}(N|\const|)$ multiplies.
Thus, GAMP requires only $\mc{O}(N\log_2 N +N|\const|)$ multiplies per 
iteration.

After the messages within the GAMP sub-graph have converged, tap-state 
beliefs are passed rightward to the MC sub-graph.
In particular, the SPA dictates that GAMP passes tap-state likelihoods
or, equivalently, the extrinsic likelihood ratios $\mc{L}_j\ext$.
Since the MC sub-graph is non-loopy, only one iteration of forward-backward 
message passing is performed,\footnote{
  Message passing on the MC factor graph is a standard procedure. 
  For details, we refer the reader to \cite{MacKay:Book:03,Bishop:Book:07}.} 
after which the resulting tap-state likelihoods are passed leftward
back to GAMP, where they are treated as tap-state priors $\vec{\lambda}$ in the next
equalizer iteration.
This interaction between the GAMP and MC sub-blocks can be recognized as
an incarnation of the structured-sparse reconstruction scheme recently proposed by 
the authors in \cite{Schniter:CISS:10}.

When the tap-state likelihoods passed between GAMP and MC have 
converged,\footnote{
  More precisely, the equalizer iterations are terminated when the mean-square 
  difference in tap-state log-likelihoods falls below a threshold
  or a maximum number of equalizer iterations has elapsed.}
the equalizer iterations are terminated and messages are passed leftward
from the GAMP block.
For this, SPA dictates that a symbol-belief propagates leftward from
the $y_i$ node with the form 
\begin{eqnarray}
  p_{s_i\leftarrow y_i}(s)
  &\propto& \int_{z} \mc{CN}(y_i;s z,\nu^w) \,\mc{CN}(z;\hat{z}_i,\nu_i^z) \\
  &=& \mc{CN}(y_i;s \hat{z}_i,|s|^2\nu_i^z+\nu^w) ,	\label{eq:p_y_to_s}
\end{eqnarray}
where $(\hat{z}_i,\nu_i^z)$ play the role of soft channel estimates. 
The SPA then implies that $p_{\mc{M}_i\leftarrow s_i}(s)=p_{s_i\leftarrow y_i}(s)$.

Next, beliefs are passed leftward from each symbol-mapping node $\mc{M}_i$ 
to the corresponding bit nodes $c_{i,m}$.  From the SPA, they take the form
\begin{eqnarray}
  \lefteqn{ p_{c_{i,m}\leftarrow\mc{M}_i}(c) }\nonumber\\
  &\propto& \sum_{k=1}^{|\const|} \sum_{\vec{c}:c_m=c} p(s\of{k}\giv\vec{c}) 
  	~ p_{\mc{M}_i\leftarrow s_i}(s\of{k}) 
	%\nonumber\\&&\mbox{}\times
	\prod_{m'\neq m} p_{c_{i,m'}\rightarrow\mc{M}_i}(c_{m'}) \nonumber\\
  &=& \sum_{k: c_m\of{k}=c} p_{\mc{M}_i\leftarrow s_i}(s\of{k})
  	\frac{\prod_{m'=1}^M p_{c_{i,m'}\rightarrow\mc{M}_i}(c_{m'}\of{k})}
	{p_{c_{i,m}\rightarrow\mc{M}_i}(c)} \\
  &=& \frac{1}{p_{c_{i,m}\rightarrow\mc{M}_i}(c)} 
  	\sum_{k: c_m\of{k}=c} p_{\mc{M}_i\leftarrow s_i}(s\of{k})
  	p_{\mc{M}_i\rightarrow s_i}(s\of{k}) 
\end{eqnarray}
for pairs $(i,m)$ that do not correspond to pilot/training bits.
(Since the pilot/training bits are known with certainty, there is no need
to update their pmfs.)

Finally, messages are passed leftward into the coding/interleaving block.
Doing so is equivalent to feeding extrinsic soft bit estimates 
to a soft-input/soft-output (SISO) decoder/deinterleaver, which 
treats them as priors.
Since SISO decoding is a well-studied topic  
\cite{MacKay:Book:03,Richardson:Book:09} and high-performance implementations
are readily available (e.g., \cite{Kozintsev:SW}), we will not elaborate 
on the details here.
It suffices to say that, once the extrinsic outputs of the SISO decoder 
have been computed, they are re-interleaved and passed rightward from the 
coding/interleaving block to begin another turbo iteration.
These turbo iterations continue until either the 
decoder detects no bit errors, the soft bit estimates have converged, or a
maximum number of iterations has elapsed.

%%%%%%%%%%%%%%%%%%%%%%%%%%%%%%%%%%%%%%%%%%%%%%%%%%%%%%%%%%%%%%%%%%%%%%%%%%%%%
\section{Numerical Results}                             \label{sec:sims}

In this section, we present numerical results that compare JCED using our
GAMP-based scheme to that using soft-input soft-output (SISO) equalizers
based on linear MMSE (LMMSE) and LASSO \cite{Tibshirani:JRSSb:96},
as well as to performance bounds based on perfect channel state information 
(CSI).

\subsection{Setup}

For all results, we used irregular LDPC codes with codeword length 
$\sim\! 10000$ and average column weight $3$,
generated (and decoded) using the publicly available software
\cite{Kozintsev:SW}, with random interleaving.
We focus on the case of $N\!=\!1024$ subcarrier OFDM with $16$-QAM 
(i.e., $M\!=\!4$) operating at a spectral efficiency of $\eta\!=\!2$ bpcu.
%Note that, in order to maintain a constant $\eta$, the code rate $R$ must
%be changed as the number of pilot/training bits change.
For bit-to-symbol mapping, we used multilevel Gray-mapping \cite{deJong:TCOM:05}, 
noting recent work \cite{Samuel:ASIL:09} that conjectures the optimality
of Gray-mapping when BICM is used with a strong code.
In some simulations, we used $\Np\!>\!0$ pilot-only subcarriers and $\Mt\!=\!0$
interspersed training bits, whereas in others we used $\Np\!=\!0$ and $\Mt\!>\!0$.
When $\Np\!>\!0$, the pilot subcarriers were placed randomly and modulated with 
(known) QAM symbols chosen uniformly at random.
When $\Mt\!>\!0$, the training bits were placed at the most significant bits (MSBs)
of uniformly spaced data-subcarriers and modulated with the bit value $1$.

Realizations of the tap vector $\vec{x}[q]$ were generated 
from IEEE~802.15.4a outdoor-NLOS impulse responses and SRRC pulses, 
as described in \secref{IEEE}, and \emph{not} from the GM2-HMM model.
The tap vectors generated for our simulations are thus as realistic as one
can hope to obtain in software.
All reported results are averaged over $5000$ channel 
realizations (i.e., $10^7$ info bits).
%Using $\SNR$ to denote the received signal-to-noise ratio 
%on each subcarrier, we note that $E_b/N_o\!=\!\SNR/\eta$.

The GM2-HMM parameters $\vec{\nu}^0,\vec{\nu}^1,\vec{p}^{01},\vec{p}^{10}$ 
were fit from $10000$ realizations of the tap-vector
$\vec{x}$ using the procedure described in \secref{IEEE}.
In doing so, we implicitly assumed\footnote{ 
  If, instead, we knew that the receiver would be used in a different
  operating scenario, then we could generate representative realizations of 
  $\vec{x}$ for that scenario and fit the GM2-HMM parameters accordingly.
  Furthermore, one could optimize the receiver for any desired balance 
  between ``typical'' and ``worst-case'' operating conditions  
  by simply choosing appropriate training realizations $\vec{x}$. } 
that the receiver is designed for the outdoor scenario, and we leverage 
the prior information made available by the extensive measurement campaign 
conducted for the IEEE 802.15.4a standard \cite{Molisch:802.15.4a}.
In all cases, we used a \emph{maximum} of $10$ turbo iterations, $5$ equalizer 
iterations, $15$ GAMP iterations, and $25$ LDPC decoder iterations, although
in most cases the iterations converged early (as described in
\secref{jced_gamp}).
%although 
%we stress that these maxima were seldom (if ever) reached.\footnote{ 
%  \textr{For example, after the second turbo iteration, it is typical to see only a 
%  single equalizer iteration and a single GAMP iteration.} }

\subsection{Comparison with other schemes}

The proposed GAMP-based equalizer was compared with soft-input soft-output 
(SISO) equalizers based on LMMSE and LASSO \cite{Tibshirani:JRSSb:96}, 
whose constructions are now detailed.

All SISO equalizers are provided with the soft inputs $\hvec{s}[q]$ and 
$\vec{\nu}^s[q]$, i.e., the means and variances, respectively, of the symbols
$\vec{s}[q]\in\const^N$.
(Note that, if certain elements in $\vec{s}[q]$ are known perfectly as pilots, 
then the corresponding elements in $\vec{\nu}^s[q]$ will be zero-valued.)
Then, writing $\vec{s}[q]=\hvec{s}[q]+\tvec{s}[q]$, where $\tvec{s}[q]$ 
an unknown zero-mean deviation, the subcarrier observations 
$\vec{y}[q]=\Diag(\vec{s}[q])\vec{\Phi}\vec{x}[q]+\vec{w}[q]$ can be written as
\begin{eqnarray}
\vec{y}[q] 
&=& \Diag(\vec{s}[q])\vec{\Phi}\vec{x}[q]+\vec{v}[q],	\label{eq:yq}
\end{eqnarray}
where
$\vec{v}[q]\defn\Diag(\tvec{s}[q])\vec{\Phi}\vec{x}[q]+\vec{w}[q]$
is a zero-mean noise.
Treating the elements within $\tvec{s}[q]$ as uncorrelated and doing the same 
with $\vec{x}[q]$, and leveraging the fact that $\vec{\Phi}$
is a truncated DFT matrix, it is straightforward to show that 
$\cov(\vec{v}[q]) = \Diag(\vec{\nu}^v[q])$ with
$\vec{\nu}^v[q] = \nu^w\vec{1} + (\vec{1}\tran\vec{\rho})\vec{\nu}^s[q]$,
where $\vec{\rho}$ denotes the channel's PDP.
Without loss of generality, \eqref{yq} can then be converted to 
the equivalent white-noise model
\begin{eqnarray}
\vec{u}[q] 
&\defn& \Diag(\vec{\nu}^v[q])^{-\frac{1}{2}} \vec{y}[q] 
= \vec{A}\vec{x}[q]+\vec{n}[q],		\label{eq:u}
\end{eqnarray}
where $\cov(\vec{n}[q])=\vec{I}$ and
$\vec{A}[q]\defn\Diag(\vec{\nu}^v[q]^{-\frac{1}{2}}\vec{s}[q])\vec{\Phi}$
is a known matrix.
In summary, \eqref{u} provides a mechanism to handle soft inputs for both 
LASSO and LMMSE.

For LMMSE equalization, we first used \eqref{u} to compute
\begin{eqnarray}
  \hvec{x}\lmmse[q]
  &=& \Diag(\vec{\rho})\vec{A}\herm[q]
  	\big(\vec{A}[q]\Diag(\vec{\rho})\vec{A}\herm[q] 
  	+ \vec{I}\big)^{-1} \vec{u}[q] 	\quad
\end{eqnarray}
from which we obtain the subcarrier gain estimate
$\hvec{z}\lmmse[q]=\vec{\Phi}\hvec{x}\lmmse[q]$.
The covariance matrix of $\hvec{z}\lmmse[q]$ is \cite{Poor:Book:94}
\begin{equation*}
  \vec{\Phi} \big(\! \Diag(\vec{\rho}) - \Diag(\vec{\rho})\vec{A}\herm[q]
  	\big(\vec{A}[q]\Diag(\vec{\rho})\vec{A}\herm[q]+\vec{I}\big)^{-1}
  	%\nonumber\\&&\mbox{}\times
	\vec{A}[q]\Diag(\vec{\rho}) \!\big) \vec{\Phi}\herm 
	%, \quad	\label{eq:mu_x_lmmse}
\end{equation*}
whose diagonal elements $\vec{\nu}^z\lmmse[q]$ are variances on the
gain estimates $\hvec{z}\lmmse[q]$.
Finally, we obtain soft symbol estimates from
the soft gain estimates $(\hvec{z}\lmmse[q],\vec{\nu}^z\lmmse[q])$ 
via \eqref{p_y_to_s}. 

For LASSO,\footnote{
  The criterion employed by LASSO \cite{Tibshirani:JRSSb:96} is equivalent
  to the one employed in ``basis pursuit denoising'' \cite{Chen:JSC:98}.}
we first computed the tap estimate $\hvec{x}\lasso[q]$ from \eqref{u} 
using the celebrated SPGL1 algorithm \cite{vandenBerg:JSC:08}.
In doing so, we needed to specify the target residual variance, i.e., 
$\nu^u\lasso\!\defn\!
\frac{1}{N}\norm{\vec{u}[q]-\vec{A}[q]\hvec{x}\lasso[q]}_2^2$. 
Because $\cov(\vec{n}[q])=\vec{I}$, we expect the optimal value of 
$\nu^u\lasso$ to be near $1$ and,
after extensive experimentation, we found that the value
$\nu^u\lasso\!=\!0.9$ works well at high SNR and that the value 
$\nu^u\lasso\!=\!1.5$ works well at low SNR. 
Thus, for each $\vec{u}[q]$, we computed SPGL1 estimates using each 
of these two\footnote{
  We also tried running SPGL1 for a dense grid of $\nu^u\lasso$
  values, but often it would get ``stuck'' at one of them and eventually
  return an error.} 
targets, and kept the one that minimized the squared error 
$\nu^x\lasso[q]\defn\frac{1}{L}\norm{\vec{x}[q]-\hvec{x}\lasso[q]}_2^2$,
which we assume a genie is able to provide.
For the soft outputs, we set 
$\hvec{z}\lasso[q]=\vec{\Phi}\hvec{x}\lasso[q]$ and take 
$\vec{\nu}^z\lasso[q]$ to be the diagonal elements of 
$\vec{\Phi}\cov(\hvec{x}\lasso[q])\vec{\Phi}\herm$.
Assuming $\cov(\hvec{x}\lasso[q])=\nu^x\lasso[q]\vec{I}$
and leveraging the fact that $\vec{\Phi}$ is a truncated DFT matrix, 
we find $\vec{\nu}^z\lasso[q]= L\nu^x\lasso[q] \vec{1}$.
Finally, using \eqref{p_y_to_s}, we obtain soft symbol estimates from
the soft gain estimates $(\hvec{z}\lasso[q],\vec{\nu}^z\lasso[q])$.
Due to the genie-aided steps, the performance attained
by our LASSO implementation is better than what could be obtained in practice.

These LMMSE- and LASSO-based SISO equalizers were then embedded in the 
overall factor graph in the same manner as GAMP, with the following 
exceptions: 
1) The LMMSE and LASSO algorithms could not be connected to the MC sub-block,
since they are not based on a two-state mixture model;
2) For LASSO, if the genie-aided MSE $\nu^x\lasso[q]$ did not improve 
during a given turbo iteration, then the corresponding outputs 
$(\hvec{z}\lasso[q],\vec{\nu}^z\lasso[q])$ were not updated.
This rule was employed to prevent turbo-LASSO from occasionally diverging 
at low SNR;
3) For LASSO, if $\Np\!>\!0$ and $\Mt\!=\!0$, then the LASSO estimates computed 
during the first turbo iteration use only pilot subcarriers.
This makes the performance of SISO-LASSO after the first turbo iteration 
equal to the performance of the standard pilot-aided LASSO.
%The performance of turbo-LMMSE attained after the first turbo iteration is
%also equal to the performance of standard (non-turbo) LMMSE.

\subsection{$\BER$ versus the number of pilot subcarriers $\Np$}

\Figref{ber_vs_Np} shows bit error rate ($\BER$) versus the number of pilot subcarriers
$\Np$ at $E_b/N_o\!=\!11$ dB and a fixed spectral efficiency of $\eta\!=\!2$ bpcu.
In this and other figures, ``\textsf{\small ALG-\#}'' refers to algorithm
\textsf{\small ALG} with \# turbo iterations (and ``\textsf{\small ALG-fin}'' 
after turbo convergence; see \figref{time_vs_snr_Np}) with the MC block 
disconnected (i.e., there was no attempt to exploit tap clustering).
Meanwhile ``\textsf{\small GAMP-\# MC-5}'' refers to GAMP+MC 
after \# turbo iterations, each containing $5$ equalizer iterations. 
Finally, \textsf{\small PCSI} refers to MAP equalization under perfect 
CSI, which yields a bound on the $\BER$ performance of any equalizer.

The curves in \figref{ber_vs_Np} exhibit a ``U'' shape because,
as $\Np$ increases, the code rate $R$ must decrease to maintain the
fixed spectral efficiency $\eta=2$ bpcu.
While an increase in $\Np$ generally makes channel estimation easier,
the reduction in $R$ makes data decoding more difficult.
For all schemes under comparison, \figref{ber_vs_Np} suggests that the
choice $\Np\!\approx\!224$ is optimal under the operating conditions.
Overall, we see GAMP significantly outperforming both LMMSE and LASSO. 
Moreover, we see a small but definite gain from the MC block.

\putFrag{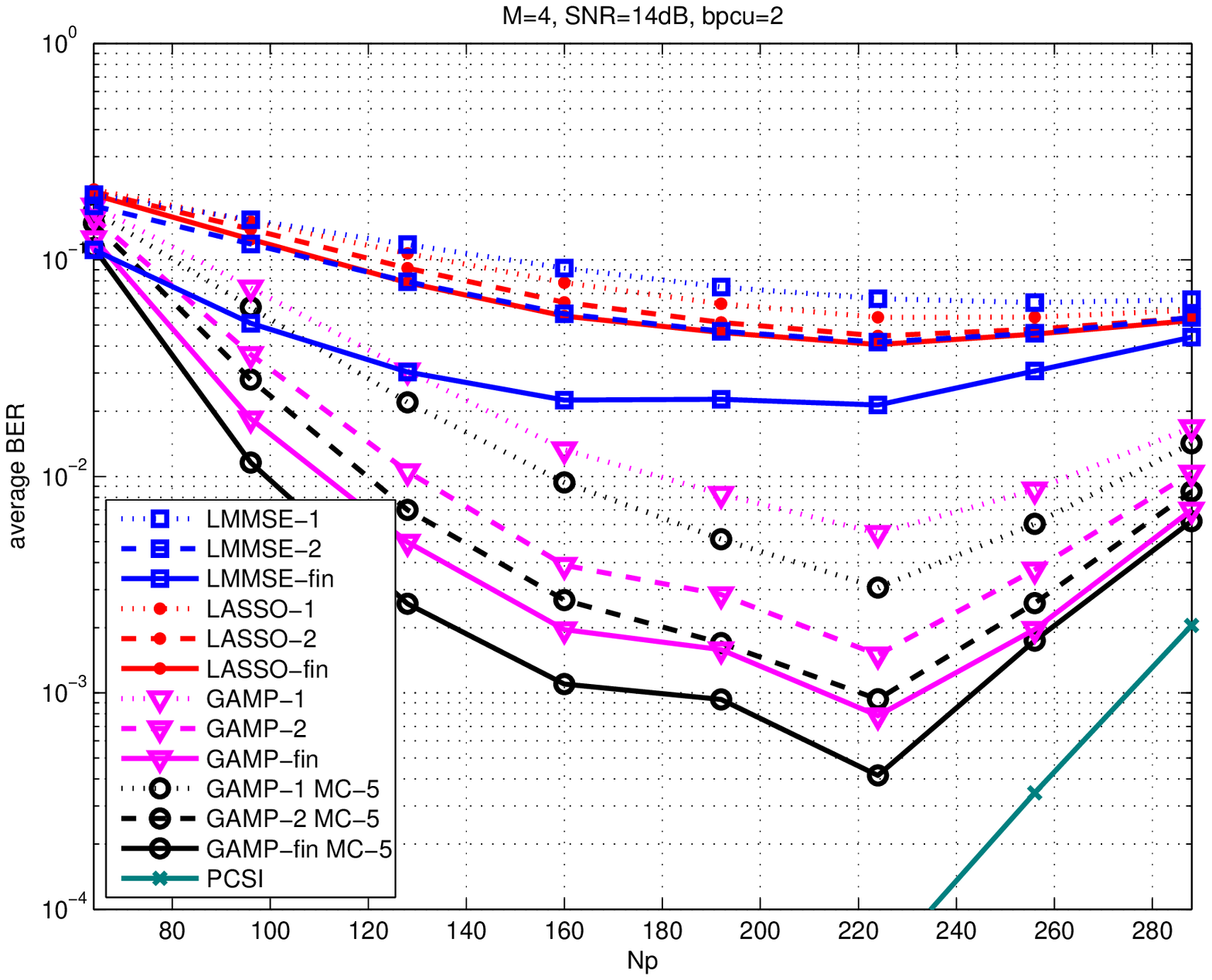}
        {$\BER$ versus number of pilot subcarriers $\Np$, for
         $E_b/N_o\!=\!11$ dB, %$\SNR\!=\!14$dB,
         $\Mt\!=\!0$ training bits,
         $\eta\!=\!2$ bpcu,
         and $16$-QAM.}
        {\figsize}
        {\psfrag{Np}[t][t][0.9]{$\Np$}
         \psfrag{average BER}[][][0.9]{$\BER$}
         \psfrag{M=4, SNR=14dB, bpcu=2}[][][0.9]{}}

\subsection{$\BER$ versus the number of interspersed training bits $\Mt$}

Although $\Np>0$ pilot subcarriers are required for decoupled channel
estimation and decoding, JCED can function with $\Np\!=\!0$ as 
long as a sufficient number $\Mt$ of training bits are interspersed among 
the coded bits used to construct each QAM symbol.
To examine this latter case, \figref{ber_vs_Mt} shows $\BER$ versus 
$\Mt$ at $E_b/N_o\!=\!10$ dB,
a fixed spectral efficiency of $\eta\!=\!2$ bpcu, and $\Np\!=\!0$.
Again we see the ``U'' shape, but with GAMP working very well for a 
relatively wide range of $\Mt$, and again 
we see a small but noticeable BER improvement when the 
MC block is used.
SISO-LASSO seems to work to some degree with $\Np=0$, 
but SISO-LASSO does not.

\putFrag{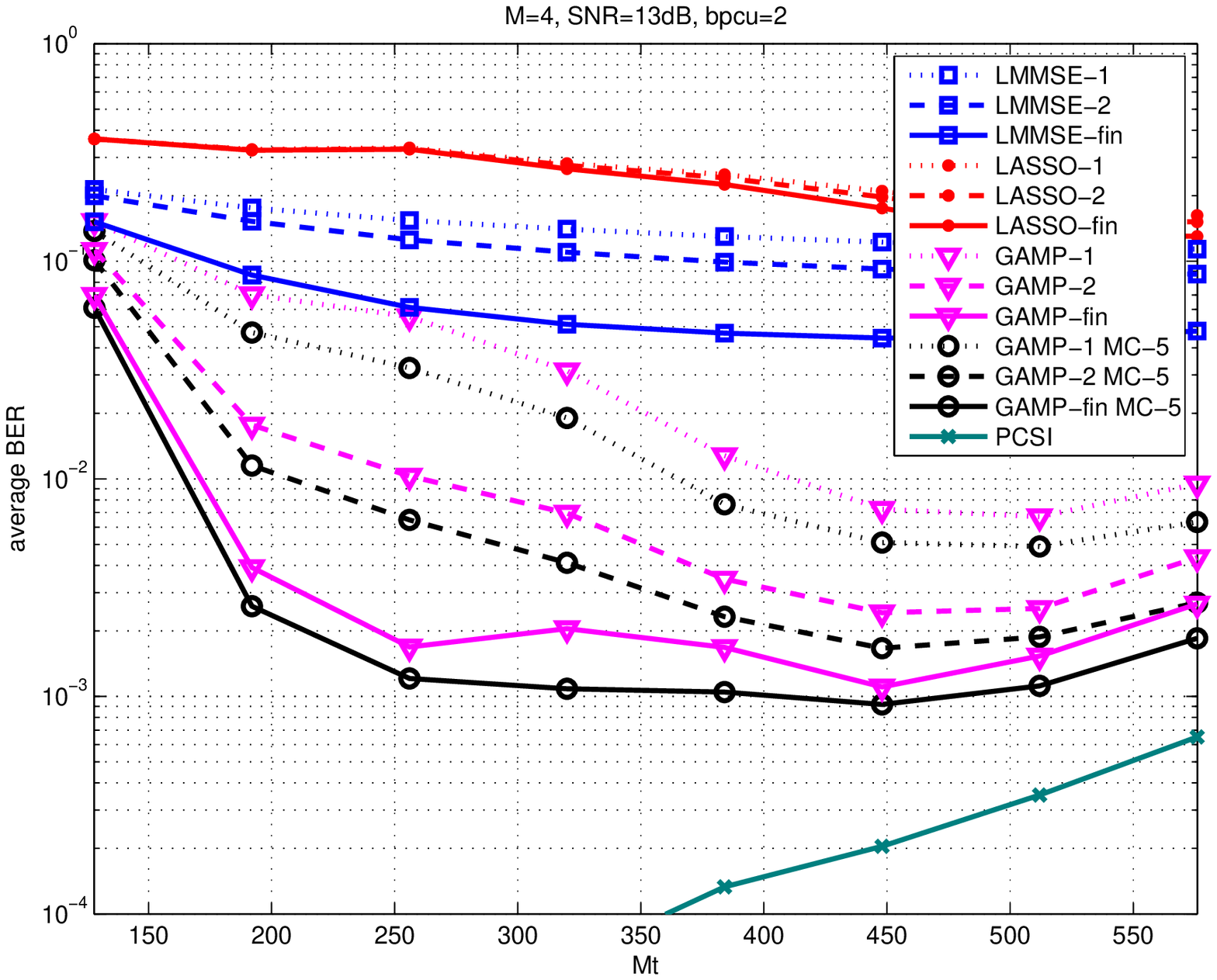}
        {$\BER$ versus number of interspersed training bits $\Mt$, for
         $E_b/N_o=10$ dB, %$\SNR\!=\!13$dB,
         $\Np\!=\!0$ pilots subcarriers,
         $\eta\!=\!2$ bpcu,
         and $16$-QAM.}
        {\figsize}
        {\psfrag{Mt}[t][t][0.9]{$\Mt$}
         \psfrag{average BER}[][][0.9]{$\BER$}
         \psfrag{M=4, SNR=13dB, bpcu=2}[][][0.9]{}}

\subsection{$\BER$ versus $E_b/N_o$}
\Figref{ber_vs_snr_Np} shows $\BER$ versus $E_b/N_o$ using 
$\Np\!=\!224$ pilot subcarriers (as suggested by \figref{ber_vs_Np})
and $\Mt\!=\!0$ training bits.
Relative to the perfect-CSI bound, we see SISO-LASSO performing within 
$5$ dB during the first turbo iteration and within $4.5$ dB 
after convergence.
Meanwhile, we see SISO-LMMSE performing very poorly during the first
turbo iteration, but eventually surpassing SISO-LASSO and coming 
within $4$ dB from the perfect-CSI bound.
Remarkably, we see GAMP+MC performing within $0.6$ dB of 
the perfect-CSI bound (and within $1$ dB after only $2$ turbo iterations).
This excellent performance confirms that the proposed GM2-HMM channel model 
and equalizer design together do an excellent job of capturing and
exploiting the lag-dependent clustered-sparse characteristics of the 
802.15.4a channel taps.
Comparing the GAMP traces to the GAMP+MC traces, we see that the MC block
yields a small but noticeable benefit. 

\Figref{ber_vs_snr_Mt} shows $\BER$ versus $E_b/N_o$ using 
$\Mt\!=\!448$ interspersed training bits (as suggested by \figref{ber_vs_Mt})
and $\Np\!=\!0$ pilot subcarriers.
There we see that SISO-LASSO does not perform well at all. 
SISO-LMMSE works to some degree after several turbo iterations,
although not as well as in the $\Np>0$ case.
Meanwhile, we see GAMP+MC performing within $1$ dB of the perfect-CSI case,
and GAMP alone performing within $1.5$ dB.
Comparing \figref{ber_vs_snr_Mt} to \figref{ber_vs_snr_Np},
we see GAMP with training bits performing about $1$ dB 
better than GAMP with dedicated pilot subcarriers.
The perfect-CSI bound likewise improves because, with $16$-QAM, $\Mt\!=\!448$ 
training bits constitutes half the overhead of $\Np\!=\!224$ pilot subcarriers,
allowing \figref{ber_vs_snr_Mt} the use of a stronger code at $\eta\!=\!2$ bpcu.
%A similar observation was made in \cite{Schniter:ASIL:10,Schniter:PHYCOM:11}
%under a simpler Bernoulli-Gaussian channel model and different message 
%passing algorithms.

\putFrag{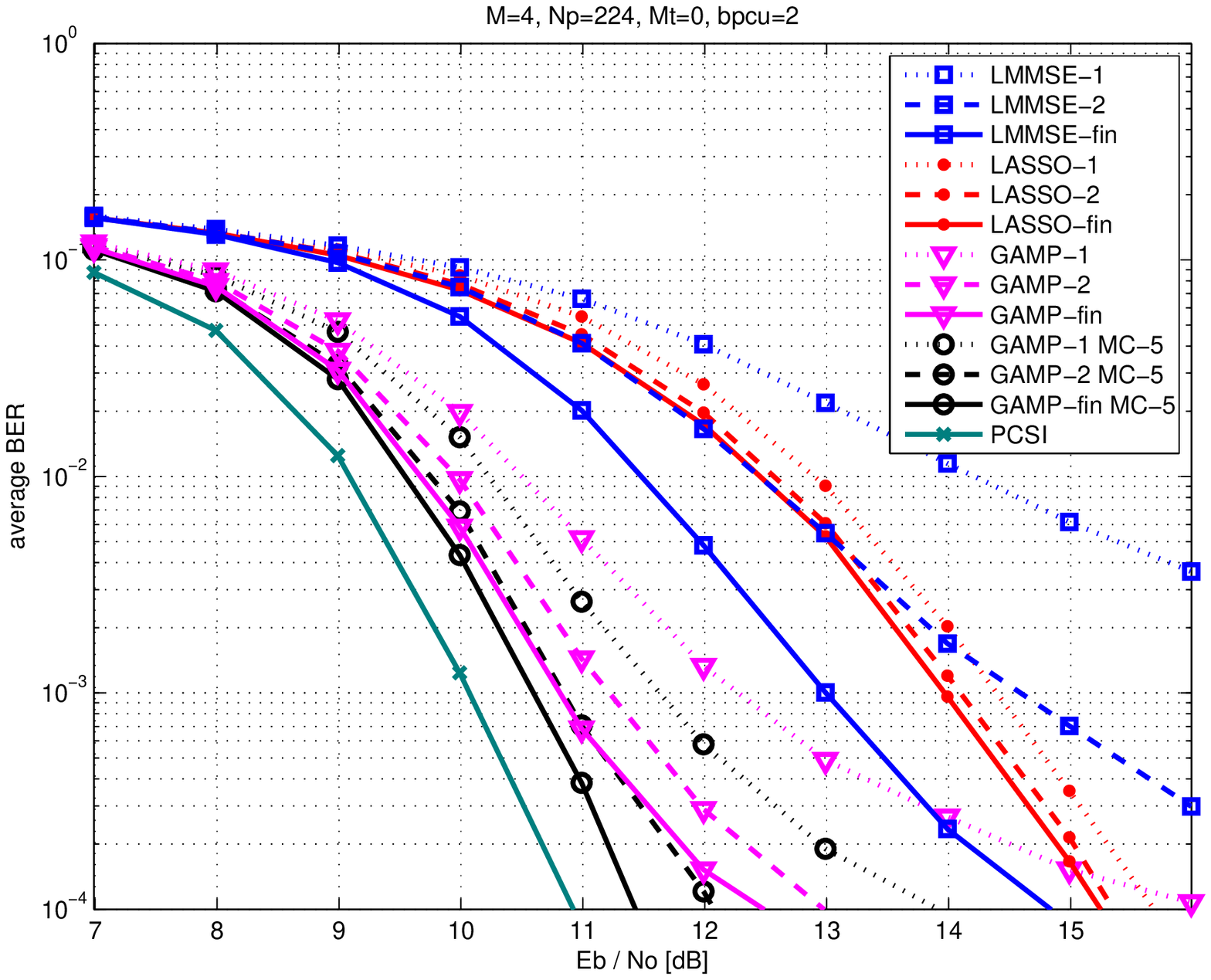}
        {$\BER$ versus $E_b/N_o$, for
         $\Np\!=\!224$ pilot subcarriers,
         $\Mt\!=\!0$ training bits,
         $\eta\!=\!2$ bpcu,
         and $16$-QAM.}
        {\figsize}
        {\psfrag{Eb / No [dB]}[][][0.8]{\sf $E_b/N_o$ [dB]}
         \psfrag{average BER}[][][0.9]{$\BER$}
         \psfrag{M=4, Np=224, Mt=0, bpcu=2}[][][0.9]{}}

\putFrag{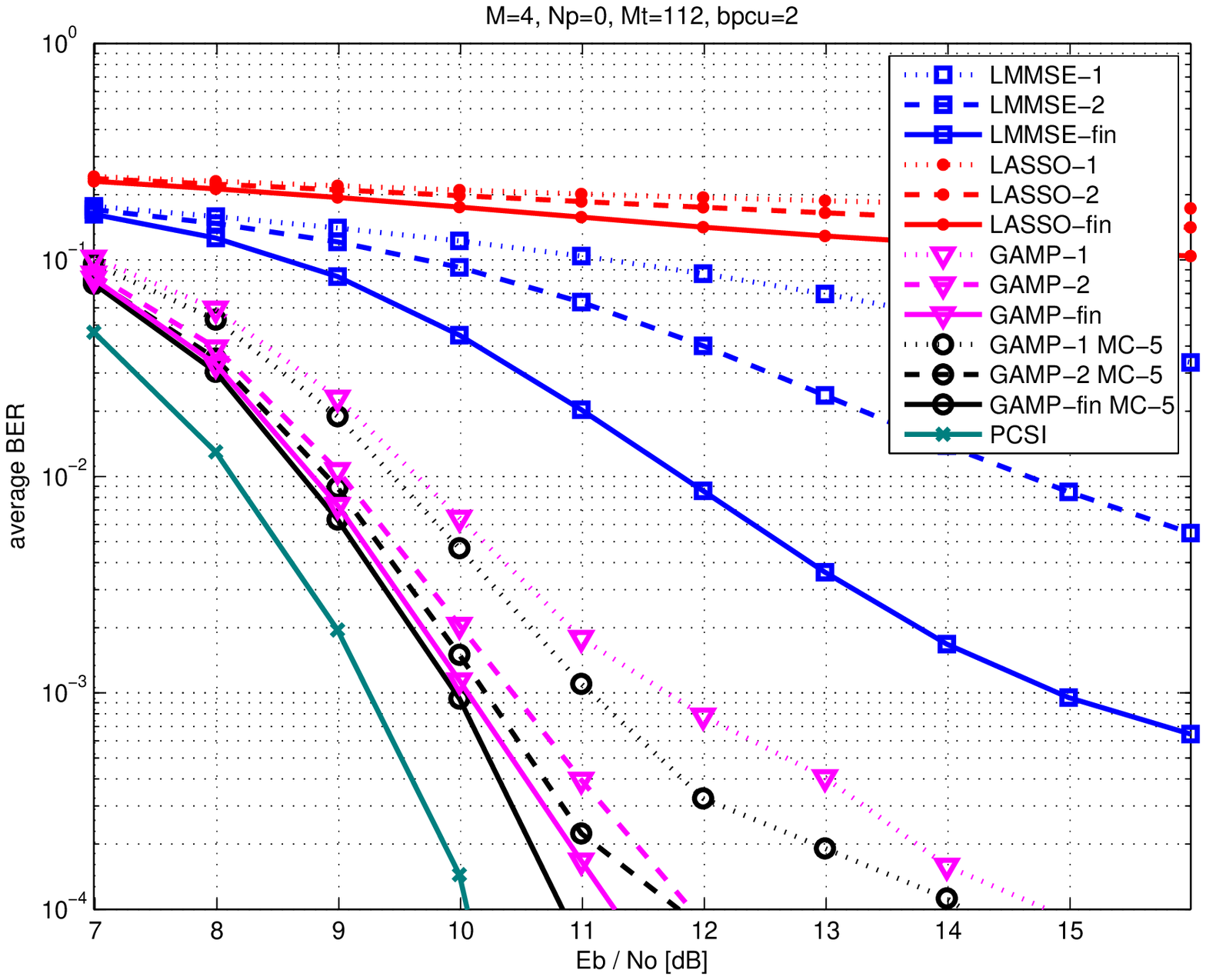}
        {$\BER$ versus $E_b/N_o$, for
         $\Np\!=\!0$ pilot subcarriers,
         $\Mt\!=\!448$ training bits,
         $\eta\!=\!2$ bpcu,
         and $16$-QAM.}
        {\figsize}
        {\psfrag{Eb / No [dB]}[][][0.8]{\sf $E_b/N_o$ [dB]}
         \psfrag{average BER}[][][0.9]{$\BER$}
         \psfrag{M=4, Np=0, Mt=112, bpcu=2}[][][0.9]{}}

\subsection{Channel-tap $\NMSE$ versus $E_b/N_o$}

\Figref{nmse_vs_snr_Np} shows the channel estimates' normalized mean-squared 
error ($\NMSE$) $\E\{\|\vec{x}[q]\!-\!\hvec{x}[q]\|_2^2/\|\vec{x}[q]\|_2^2\}$
versus $E_b/N_o$, at the point that the turbo iterations were terminated, 
using $\Np\!=\!224$ pilot subcarriers and $\Mt\!=\!0$ training bits.
(For comparison, \figref{ber_vs_snr_Np} shows $\BER$ for this configuration.)
We also show the $\NMSE$ attained by the ``bit and support
genie'' (BSG), which calculates MMSE channel estimates using
perfect knowledge of both the coded bits and the hidden channel states 
$\{d_j\}$, and which provides a lower bound for any channel estimator.
In the figure, we see that the $\NMSE$s of LMMSE and LASSO channel estimates 
are within $8$-to-$12$ dB of the BSG, whereas those of GAMP are within 
$2$-to-$4$ dB.
Meanwhile, we see that GAMP+MC has a small but noticeable advantage over
GAMP alone.
We reason that the LMMSE estimates are worse than the GAMP estimates because 
they do not exploit the non-Gaussianity of the channel taps $x_j$, and the
LASSO estimates are worse than the GAMP estimates because they do not
exploit the known priors on the channel taps (i.e., the
lag-dependent sparsity $\vec{\lambda}$ and PDP $\vec{\rho}$).

\putFrag{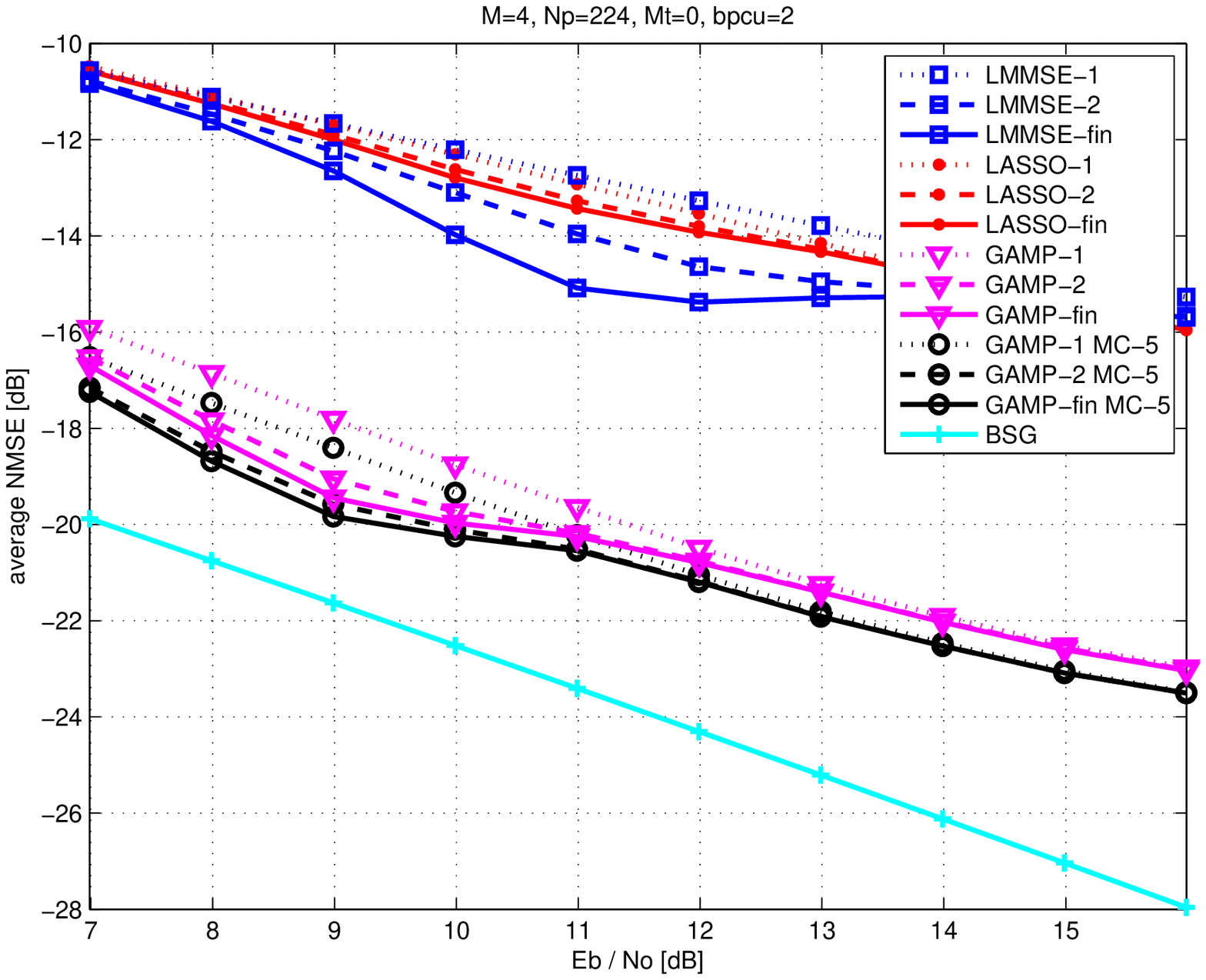}
        {Channel tap $\NMSE$ versus $E_b/N_o$, for
         $\Np\!=\!224$ pilot subcarriers,
         $\Mt\!=\!0$ training bits,
         $\eta\!=\!2$ bpcu,
         and $16$-QAM.}
        {\figsize}
        {\psfrag{Eb / No [dB]}[][][0.8]{\sf $E_b/N_o$ [dB]}
         \psfrag{average NMSE [dB]}[][][0.9]{$\NMSE$ \sf \small [dB]}
         \psfrag{M=4, Np=224, Mt=0, bpcu=2}[][][0.9]{}}

\subsection{Computational complexity versus $E_b/N_o$}

\Figref{time_vs_snr_Np} shows the average time per turbo iteration
(in Matlab seconds on a $2.6$GHz CPU), the average number of turbo 
iterations, and the average total time (to turbo convergence), as a function 
of $E_b/N_o$, using $\Np\!=\!224$ pilot subcarriers and $\Mt\!=\!0$ training 
bits.
(For comparison, \figref{ber_vs_snr_Np} shows $\BER$ for this configuration
and \figref{nmse_vs_snr_Np} shows $\NMSE$.)
Regarding the average time per turbo iteration, we see GAMP$\pm$MC
taking $\approx 1.5$ sec at low $E_b/N_o$ and $\approx 0.5$ sec at 
high $E_b/N_o$.
GAMP+MC takes only slightly longer than GAMP alone due to the efficiency of 
the message computations within the MC block, and the fact that both the
GAMP iterations and equalizer iterations are terminated as soon as the
messages converge.  
In comparison, SISO-LMMSE takes $\approx 4.5$ sec per turbo iteration, and 
SISO-LASSO takes between $1$ and $7$ sec, depending on $E_b/N_o$.
Regarding the number of average number of turbo iterations until convergence,
we see that---at low $E_b/N_o$---% 
GAMP+MC takes about $5$ turbo iterations,
GAMP alone takes about $7$, 
SISO-LMMSE takes about $5$, and
SISO-LASSO takes about $3$, while---at high $E_b/N_o$---%
all algorithms converge after only $1$ turbo iteration.
Regarding the total time for equalization, GAMP+MC and GAMP are about
the same at low $E_b/N_o$, whereas GAMP alone takes about $30\%$ less time 
at high $E_b/N_o$.
Meanwhile, SISO-LASSO and SISO-LMMSE are uniformly slower than GAMP and GAMP+MC over the entire $E_b/N_o$ range, in some cases by a factor of $10$.

\putFrag{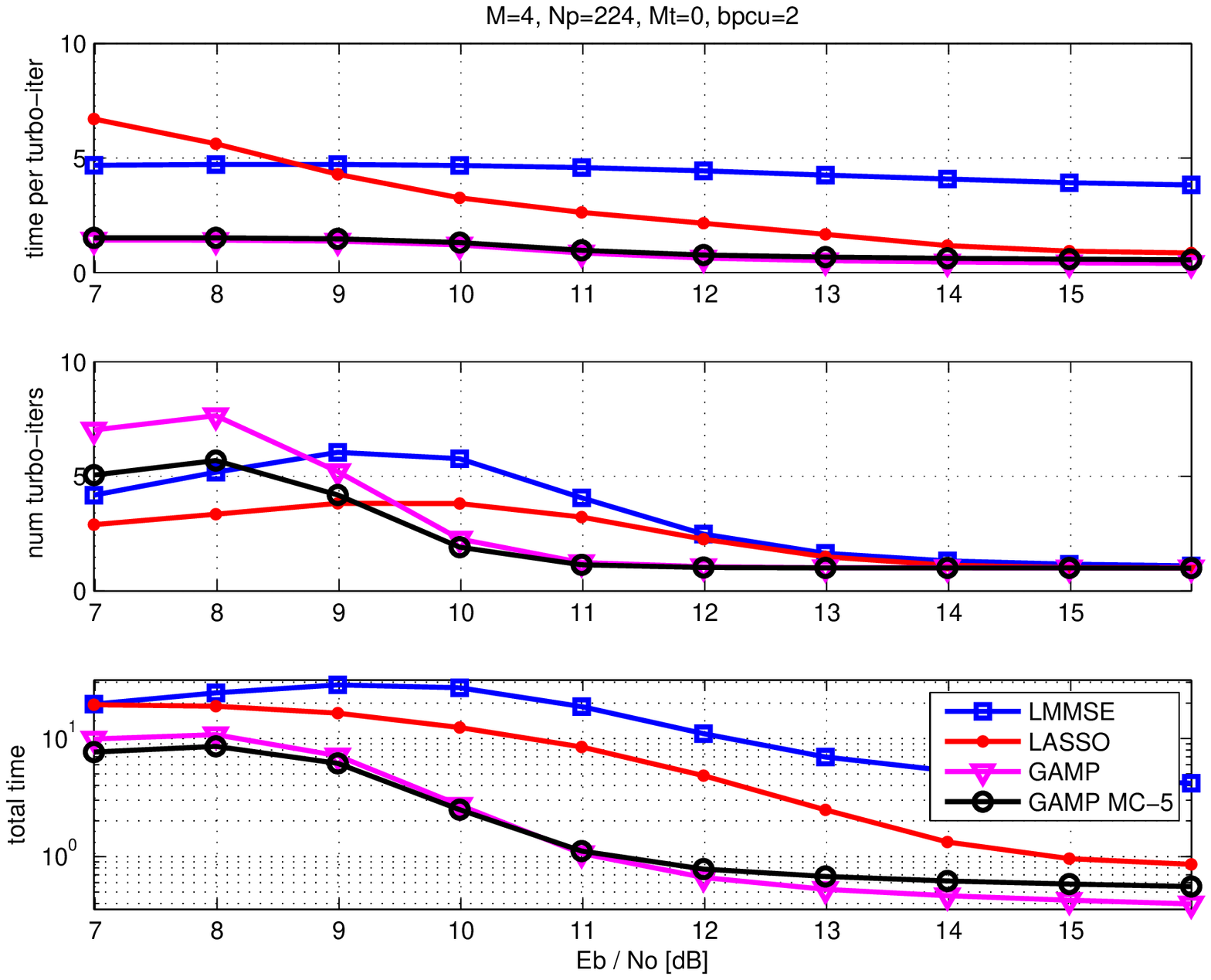}
        {Average time per turbo iteration (top),
	 average number of turbo iterations (middle), and 
	 average total time (bottom), versus $E_b/N_o$, for
         $\Np\!=\!224$ pilot subcarriers,
         $\Mt\!=\!0$ training bits,
         $\eta\!=\!2$ bpcu,
         and $16$-QAM.}
        {\figsize}
        {\psfrag{Eb / No [dB]}[][][0.8]{\sf $E_b/N_o$ [dB]}
         \psfrag{total time}[][][0.7]{\sf total time}
         \psfrag{num turbo-iters}[b][][0.7]{\sf \# turbo iters}
         \psfrag{time per turbo-iter}[b][][0.65]{\sf time per turbo iter}
         \psfrag{M=4, Np=224, Mt=0, bpcu=2}[][][0.9]{}}

%\subsection{$\BER$ under an i.i.d Bernoulli-Gaussian channel model}
%
%\putFrag{ber_vs_snr_Mt_iid}
%        {$\BER$ versus $E_b/N_o$, for
%         $\Np\!=\!0$ pilot subcarriers,
%         $\Mt\!=\!448$ training bits,
%         $\eta\!=\!2$ bpcu,
%         $16$-QAM, and i.i.d-BG equalization.}
%        {\figsize}
%        {\psfrag{average BER}[][][0.9]{$\BER$}
%         \psfrag{Eb / No [dB]}[][][0.8]{\sf $E_b/N_o$ [dB]}
%         \psfrag{M=4, Np=0, Mt=464, bpcu=2}[][][0.9]{}}
%
%%%%%%%%%%%%%%%%%%%%%%%%%%%%%%%%%%%%%%%%%%%%%%%%%%%%%%%%%%%%%%%%%%%%%%%%%%%%_
\section{Conclusion}				\label{sec:conc}

In this paper, we presented a factor-graph approach to joint channel-estimation
and decoding (JCED) for BICM-OFDM that merges recent advances in 
approximate message passing algorithms \cite{Rangan:10b} with those in 
structured-sparse signal reconstruction \cite{Schniter:CISS:10} and 
SISO decoding \cite{MacKay:Book:03}.
Different from existing factor-graph approaches to JCED, ours is able to 
exploit not only sparse channel taps, but also clustered sparsity patterns
that typify large-bandwidth communication channels, such as those that result 
from pulse-shaped communication over IEEE~802.15.4a modeled channels.
For this purpose, we proposed the use of a two-state Gaussian mixture prior 
with a Markov model on the hidden tap states.
The implementation complexity of our JCED scheme is dominated by 
$\mc{O}(N\log_2 N \!+\! N|\const|)$ multiplies per GAMP iteration, 
facilitating the application to systems with many subcarriers $N$ and many
channel taps $L<N$.
Experiments with IEEE~802.15.4a modeled channels showed $\BER$ performance
within $1$ dB of the known-channel bound, and $3$--$4$ dB better than 
LMMSE- and LASSO-based soft equalizers.
These experiments also suggested that, with our proposed approach, the use 
of interspersed training bits is more efficient than the use of 
dedicated pilot subcarriers.
For very large constellations (e.g., $|\const|\!=\!1024$), future work is
motivated to reduce the linear complexity dependence on $|\const|$.

%%%%%%%%%%%%%%%%%%%%%%%%%%%%%%%%%%%%%%%%%%%%%%%%%%%%%%%%%%%%%%%%%%%%%%%%%%%%%

\appendices

\section{Derivation of GAMP Functions \textnormal{$g\out$} and \textnormal{$g'\out$}}
\label{app:out}

In this appendix, we derive the GAMP quantities $g\out(y,\hat{z},\nu^z)$
and $g'\out(y,\hat{z},\nu^z)$ given in \eqref{gout}-\eqref{ei}.

From (D1), we have that
\begin{eqnarray}
  %g\out(y,\hat{z},\nu^z)
  \E_{Z_i|Y_i}\{z \giv y;\hat{z},\nu^z\}
  &=& \frac{1}{p_{Y_i}(y)} \int_z z \, p_{Y_i|Z_i}(y|z) 
        \, \mc{CN}(z;\hat{z},\nu^z)  ,  \quad           \label{eq:gout2}
\end{eqnarray}
where
$p_{Y_i}(y) \defn \int_z p_{Y_i|Z_i}(y|z) \mc{CN}(z;\hat{z},\nu^z)$.
From \eqref{pY|Z}, we rewrite $p_{Y_i|Z_i}(y|z)$ as
\begin{eqnarray}
  p_{Y_i|Z_i}(y|z)
  &=& \sum_{k=1}^{2^M} \frac{\beta_i\of{k}}{s\of{k}} 
        \,\mc{CN}\Big(z;\frac{y}{s\of{k}},\frac{\nu^w}{|s\of{k}|^2}\Big) ,
\end{eqnarray}
so that
\begin{eqnarray}
%\lefteqn{ 
  \int_z z\, p_{Y_i|Z_i}(y|z) \mc{CN}(z;\hat{z},\nu^z) 
%}\nonumber\\
  &=& \sum_{k=1}^{2^M} \frac{\beta_i\of{k}}{s\of{k}} \int_z z \,
        \mc{CN}\Big(z;\frac{y}{s\of{k}},\frac{\nu^w}{|s\of{k}|^2}\Big) 
        \mc{CN}(z;\hat{z},\nu^z) \quad \\
%\lefteqn{ 
  p_{Y_i}(y) 
%}\nonumber\\
  &=& \sum_{k=1}^{2^M} \frac{\beta_i\of{k}}{s\of{k}} \int_z 
        \mc{CN}\Big(z;\frac{y}{s\of{k}},\frac{\nu^w}{|s\of{k}|^2}\Big) 
        \mc{CN}(z;\hat{z},\nu^z).
\end{eqnarray}
Using the property that
\begin{eqnarray}
%\lefteqn{
  \mc{CN}(x;\hat{\theta},\nu^\theta)\mc{CN}(x;\hat{\phi},\nu^\phi)
%}\nonumber\\
  &=& \mc{CN}\Big(x;\frac{\hat{\theta}/\nu^\theta+\hat{\phi}/\nu^\phi}
        {1/\nu^\theta+1/\nu^\phi},\frac{1}{1/\nu^\theta+1/\nu^\phi}\Big)
%\nonumber\\&&\mbox{}\times 
        \mc{CN}(0;\hat{\theta}-\hat{\phi},\nu^\theta+\nu^\phi) ,        
                                                        \label{eq:pogr}
\end{eqnarray}
we can rewrite
\begin{eqnarray}
  \lefteqn{ 
  \int_z z \, p_{Y_i|Z_i}(y|z) \, \mc{CN}(z;\hat{z},\nu^z) 
  }\nonumber\\
  &=& \sum_{k=1}^{2^M} \frac{\beta_i\of{k}}{s\of{k}} 
        \mc{CN}\Big(0;\frac{y_i}{s}-\hat{z},\frac{\nu^w}{|s\of{k}|^2}+\nu^z\Big)
%\nonumber\\&&\mbox{}\times
        \int_z z \,
        \mc{CN}\bigg(z;\frac{
          \frac{y}{s\of{k}}\frac{|s\of{k}|^2}{\nu^w}+\frac{\hat{z}}{\nu^z}
        }{
          \frac{|s\of{k}|^2}{\nu^w}+\frac{1}{\nu^z}
        },\frac{1}{\frac{|s\of{k}|^2}{\nu^w}+\frac{1}{\nu^z}}\bigg)  
                                                        \label{eq:prod}\\
  &=& \sum_{k=1}^{2^M} \frac{\beta_i\of{k}}{s\of{k}} 
        \mc{CN}\Big(\frac{y_i}{s};\hat{z},\frac{\nu^w}{|s\of{k}|^2}+\nu^z\Big)
        \frac{
          \frac{y}{s\of{k}}\frac{|s\of{k}|^2}{\nu^w}+\frac{\hat{z}}{\nu^z}
        }{
          \frac{|s\of{k}|^2}{\nu^w}+\frac{1}{\nu^z}
        } \quad \\
  &=& \sum_{k=1}^{2^M} \beta_i\of{k}
        \mc{CN}\big(y_i;s\of{k}\hat{z},|s\of{k}|^2\nu^z+\nu^w\big)
%\nonumber\\&&\mbox{}\times
        \bigg( 
        \underbrace{ \Big(\frac{y}{s\of{k}}-\hat{z}\Big) 
                \frac{|s\of{k}|^2 \nu^z}{|s\of{k}|^2 \nu^z + \nu^w} 
        }_{\displaystyle \defn \hat{e}\of{k}(y,\hat{z},\nu^z)}
        + \hat{z} \bigg)                                \label{eq:intz}
\end{eqnarray}
and, using the same procedure, we get
\begin{eqnarray}
  p_{Y_i}(y) 
  &=& \sum_{k=1}^{2^M} \beta_i\of{k}
        \mc{CN}\big(y_i;s\of{k}\hat{z},|s\of{k}|^2\nu^z+\nu^w\big) .
                                                        \label{eq:pY}
\end{eqnarray}
With $\xi_i\of{k}(y,\hat{z},\nu^z)$ defined in \eqref{xi},
equations \eqref{gout2} and \eqref{intz} and \eqref{pY} combine to give
\begin{eqnarray}
  \E_{Z_i|Y_i}\{z \giv y;\hat{z},\nu^z\}
  &=& \sum_{k=1}^{2^M} \xi_i\of{k}(y,\hat{z},\nu^z)
        \big( \hat{e}\of{k}(y,\hat{z},\nu^z) + \hat{z} \big) .  \quad
							\label{eq:EZY}
\end{eqnarray}
Finally, from \eqref{EZY} and the definition of $g\out(y,\hat{z},\nu^z)$ 
in (D2), equation \eqref{gout} follows immediately.

From (D1), we have that
\begin{eqnarray}
%\lefteqn{ 
  \var_{Z_i|Y_i}\{z \giv y;\hat{z},\nu^z\}
%}\nonumber\\
  &=& \frac{1}{p_{Y_i}(y)} \int_z |z-\E_{Z_i|Y_i}\{z \giv y;\hat{z},\nu^z\}|^2 \, p_{Y_i|Z_i}(y|z) 
        \, \mc{CN}(z;\hat{z},\nu^z) .   \label{eq:varZY2}
\end{eqnarray}
Similar to \eqref{prod}, we can write
\begin{eqnarray}
  \lefteqn{ \int_z |z-\E_{Z_i|Y_i}\{z \giv y;\hat{z},\nu^z\}|^2 \, p_{Y_i|Z_i}(y|z) 
        \, \mc{CN}(z;\hat{z},\nu^z) }\nonumber\\
  &=& \sum_{k=1}^{2^M} \frac{\beta_i\of{k}}{s\of{k}} 
        \mc{CN}\Big(0;\frac{y_i}{s}-\hat{z},\frac{\nu^w}{|s\of{k}|^2}+\nu^z\Big)
      \nonumber\\&&\mbox{}\times
        \int_z |z-\E_{Z_i|Y_i}\{z \giv y;\hat{z},\nu^z\}|^2 \,
        \mc{CN}\bigg(z;\frac{
          \frac{y}{s\of{k}}\frac{|s\of{k}|^2}{\nu^w}+\frac{\hat{z}}{\nu^z}
        }{
          \frac{|s\of{k}|^2}{\nu^w}+\frac{1}{\nu^z}
        },\frac{1}{\frac{|s\of{k}|^2}{\nu^w}+\frac{1}{\nu^z}}\bigg) .
\end{eqnarray}
Then, using the change-of-variable $\tilde{z}\defn z-\E_{Z_i|Y_i}\{z \giv y;\hat{z},\nu^z\}$, 
and absorbing the $s\of{k}$ terms as done in \eqref{intz}, we get
\begin{eqnarray}
  \lefteqn{ \int_z |z-\E_{Z_i|Y_i}\{z \giv y;\hat{z},\nu^z\}|^2 \, p_{Y_i|Z_i}(y|z) 
        \, \mc{CN}(z;\hat{z},\nu^z) }\nonumber\\
  &=& \sum_{k=1}^{2^M} \beta_i\of{k}
        \mc{CN}\big(y_i;s\of{k}\hat{z},|s\of{k}|^2\nu^z+\nu^w\big)
      \nonumber\\&&\mbox{}\times
        \int_{\tilde{z}} |\tilde{z}|^2 \,
        \mc{CN}\Big(\tilde{z}; \hat{e}\of{k}+
        \underbrace{\hat{z}-\E_{Z_i|Y_i}\{z \giv y;\hat{z},\nu^z\}}_{\displaystyle = -\hat{e}_i},
        \frac{\nu^w\nu^z}{|s\of{k}|^2\nu^z+\nu^w}\Big)  \\
  &=& \sum_{k=1}^{2^M} \beta_i\of{k}
        \mc{CN}\big(y_i;s\of{k}\hat{z},|s\of{k}|^2\nu^z+\nu^w\big)
%\nonumber\\&&\mbox{}\times
        \Big( |\hat{e}\of{k}-\hat{e}_i|^2
        + \frac{\nu^w\nu^z}{|s\of{k}|^2\nu^z+\nu^w} \Big).\label{eq:intz2}
\end{eqnarray}
Using $\xi_i\of{k}(y,\hat{z},\nu^z)$ defined in \eqref{xi} 
and $\zeta\of{k}(y,\hat{z},\nu^z)$ defined in \eqref{zeta},
equations \eqref{pY} and \eqref{varZY2} and \eqref{intz2} combine to give
\begin{eqnarray}
  \var_{Z_i|Y_i}\{z \giv y;\hat{z},\nu^z\}
  &=& \sum_{k=1}^{2^M} \xi_i\of{k}(y,\hat{z},\nu^z) \Big( 
	%\frac{\nu^z\nu^w}{|s\of{k}|^2 \nu^z+\nu^w} 
	\frac{\nu^w \zeta\of{k}(y,\hat{z},\nu^z)}{|s\of{k}|^2} 
%\nonumber\\&&\mbox{}
	+ \big|\hat{e}_i(y,\hat{z},\nu^z)-\hat{e}\of{k}(y,\hat{z},\nu^z)\big|^2
	\Big). %\\[-9mm] \nonumber			
							\label{eq:varZY}
\end{eqnarray}
which is rewritten as $\nu^e_i(y,\hat{z},\nu^z)
\defn \var_{Z_i|Y_i}\{z \giv y;\hat{z},\nu^z\}$ in \eqref{muei}.
Finally, plugging $\nu^e_i(y,\hat{z},\nu^z)$ into the definition of 
$g'\out(y,\hat{z},\nu^z)$ in (D3), we immediately obtain \eqref{g'out}.

%%%%%%%%%%%%%%%%%%%%%%%%%%%%%%%%%%%%%%%%%%%%%%%%%%%%%%%%%%%%%%%%%%%%%%%%%%%%%%%%
\section{Derivation of GAMP Functions \textnormal{$g\inp$} and \textnormal{$g'\inp$}}
\label{app:in}

In this appendix, we derive the GAMP quantities $g\inp(\hat{r},\nu^r)$
and $g'\inp(\hat{r},\nu^r)$ given in \eqref{gin}-\eqref{alfj}.

From (D4)-(D6), we note that $g\inp(\hat{r},\nu^r)$
and $\nu^r g'\inp(\hat{r},\nu^r)$ are the mean and
variance, respectively, of the pdf
\begin{eqnarray}
  \frac{1}{Z_j} p_{X_j}\!(r) \,\mc{CN}(r;\hat{r},\nu^r) ,\label{eq:pdfin}
\end{eqnarray}
where $Z_j\defn\int_r p_{X_j}\!(r) \,\mc{CN}(r;\hat{r},\nu^r)$.
Using \eqref{pogr} together with the definition of $p_{X_j}\!(.)$
from \eqref{pxj}, we find 
\begin{eqnarray}
  \lefteqn{
  p_{X_j}\!(r) \,\mc{CN}(r;\hat{r},\nu^r)
  }\nonumber\\
  &=& \lambda_j \mc{CN}(r;0,\nu_j^1) \,\mc{CN}(r;\hat{r},\nu^r) 
%\nonumber\\&&\mbox{}
        + (1-\lambda_j) \mc{CN}(r;0,\nu_j^0) \,\mc{CN}(r;\hat{r},\nu^r) \\
% &=& \textstyle 
%       \lambda_j \mc{CN}(0;-\hat{r},\nu^1_j+\nu^r) \,
%       \mc{CN}\big(r;\frac{\hat{r}/\nu^r}{1/\nu^1_j+1/\nu^r},
%               \frac{1}{1/\nu^1_j+1/\nu^r}\big) 
%       \nonumber\\&&\mbox{} \textstyle 
%       + (1-\lambda_j) \mc{CN}(0;-\hat{r},\nu^0_j+\nu^r) \,
%       \mc{CN}\big(r;\frac{\hat{r}/\nu^r}{1/\nu^0_j+1/\nu^r},
%               \frac{1}{1/\nu^0_j+1/\nu^r}\big) \\
% &=& \textstyle 
%       \lambda_j \mc{CN}(\hat{r};0,\nu^1_j+\nu^r) \,
%       \mc{CN}\big(r;\frac{\hat{r}}{\nu^r}\frac{\nu^r\nu^1_j}{\nu^r+\nu^1_j},
%               \frac{\nu^r\nu^1_j}{\nu^r+\nu^1_j}\big) 
%       \nonumber\\&&\mbox{} \textstyle
%       + (1-\lambda_j) \mc{CN}(\hat{r};0,\nu^0_j+\nu^r) \,
%       \mc{CN}\big(r;\frac{\hat{r}}{\nu^r}\frac{\nu^r\nu^0_j}{\nu^r+\nu^0_j},
%               \frac{\nu^r\nu^0_j}{\nu^r+\nu^0_j}\big) ,
% &=& \textstyle 
%       \lambda_j \mc{CN}(\hat{r};0,\nu^1_j+\nu^r) \,
%       \mc{CN}\big(r;\hat{r}\frac{\nu^1_j}{\nu^r+\nu^1_j},
%               \nu^r\frac{\nu^1_j}{\nu^r+\nu^1_j}\big) 
%       \nonumber\\&&\mbox{} \textstyle
%       + (1-\lambda_j) \mc{CN}(\hat{r};0,\nu^0_j+\nu^r) \,
%       \mc{CN}\big(r;\hat{r}\frac{\nu^0_j}{\nu^r+\nu^0_j},
%               \nu^r\frac{\nu^0_j}{\nu^r+\nu^0_j}\big) \\
  &=& \textstyle 
        \lambda_j \mc{CN}(\hat{r};0,\nu^1_j+\nu^r) \,
        \mc{CN}\big(r;\hat{r}\gamma^1_j(\nu^r),\nu^r\gamma^1_j(\nu^r)\big) 
        \nonumber\\&&\mbox{} \textstyle
        + (1-\lambda_j) \mc{CN}(\hat{r};0,\nu^0_j+\nu^r) \,
        \mc{CN}\big(r;\hat{r}\gamma^0_j(\nu^r),\nu^r\gamma^0_j(\nu^r)\big) 
\end{eqnarray}
for $\gamma^0_j(\nu^r) \defn (1+\nu^r/\nu_j^0)^{-1}$ and
$\gamma^1_j(\nu^r) \defn (1+\nu^r/\nu_j^1)^{-1}$.
%for $\gamma^0_j(\nu^r)$ and $\gamma^1_j(\nu^r)$ defined in \eqref{gam0j}-\eqref{gam1j}.
This implies that
\begin{eqnarray}
  Z_j
  &=& \lambda_j \mc{CN}(\hat{r};0,\nu^1_j+\nu^r)
        + (1-\lambda_j) \mc{CN}(\hat{r};0,\nu^0+\nu^r) .
\end{eqnarray}
Thus, the mean obeys
\begin{eqnarray}
%\lefteqn{
  g\inp(\hat{r},\nu^r)
%}\nonumber\\
  &=& \frac{1}{Z_j}\int_r r \, p_{X_j}\!(r) \,\mc{CN}(r;\hat{r},\nu^r) \\
  &=& \underbrace{ \frac{\lambda_j\mc{CN}(\hat{r};0,\nu_j^1+\nu^r)}{Z_j} 
        }_{\displaystyle = \alpha_j(\hat{r},\nu^r)}
	\gamma_j^1(\nu^r) \,\hat{r} 
  	+ \underbrace{ \frac{(1-\lambda_j)\mc{CN}(\hat{r};0,\nu_j^0+\nu^r)}{Z_j} 
        }_{\displaystyle = 1-\alpha_j(\hat{r},\nu^r)}
	\gamma_j^0(\nu^r) \,\hat{r} ,
                                                        \label{eq:gin2}
% &=& \frac{\lambda_j \mc{CN}(\hat{r};0,\nu_j+\nu^r) 
%       \frac{\hat{r}}{\nu^r}\frac{\nu^r\nu_j}{\nu^r+\nu_j}}
%       {\lambda_j \mc{CN}(\hat{r};0,\nu_j+\nu^r)
%       + (1-\lambda_j) \mc{CN}(\hat{r};0,\nu^r)} .     \label{eq:gin2}
\end{eqnarray}
yielding \eqref{gin}, where a straightforward manipulation relates
the expression for $\alpha_j(\hat{r},\nu^r)$ above with its
definition in \eqref{alfj}.

Since, for the pdf in \eqref{pdfin}, $g\inp$ is the mean and
$\nu^r g'\inp$ is the variance, we can write
\begin{eqnarray}
%\lefteqn{
  \nu^r g'\inp(\hat{r},\nu^r)
%}\nonumber\\
% &=& \frac{1}{Z_j}\int_r |r-g\inp|^2 \, p_{X_j}\!(r) 
%       \,\mc{CN}(r;\hat{r},\nu^r)  \\
  &=& \frac{1}{Z_j}\int_r |r|^2 \, p_{X_j}\!(r) 
        \,\mc{CN}(r;\hat{r},\nu^r)  -|g\inp|^2  \\
  &=& \alpha_j \big( |\hat{r}\gamma^1_j|^2 + \nu^r\gamma^1_j \big)
   	+(1-\alpha_j) \big( |\hat{r}\gamma^0_j|^2 + \nu^r\gamma^0_j \big)
	%\nonumber\\&&\mbox{}
        -\big|\alpha_j\gamma^1_j\hat{r} + (1-\alpha_j)\gamma^0_j\hat{r}\big|^2 ,
						\label{eq:g'in2}
\end{eqnarray}
which can be simplified to yield \eqref{g'in}.

%%%%%%%%%%%%%%%%%%%%%%%%%%%%%%%%%%%%%%%%%%%%%%%%%%%%%%%%%%%%%%%%%%%%%%%%%%%%%
\bibliographystyle{ieeetr}
\bibliography{macros_abbrev,books,misc,comm,multicarrier,sparse,stc,underwater}
\def\baselinestretch{1.0}
\end{document}